\documentclass[  twocolumn]{aastex701}

\usepackage{lipsum}
\usepackage{longtable}
\usepackage{comment}
\usepackage{float}
\usepackage{chemformula}
\usepackage{placeins}

\newcommand{\pcms}{cm$^{-2}$}

\newcommand{\kms}{\mbox{km s$^{-1}$}}

\newcommand\e[1]{\ensuremath{ 10^{#1}}}

\begin{document}

\title{Non-detection of HC(S)SH: Estimating Upper Limits and Constraining Chemistry\footnote{Refuting the initial reported detection of HC(S)SH towards IRAS 4A2.}}

\correspondingauthor{Dipen Sahu}
\email{dsahu@prl.res.in}

\author[0000-0002-4393-3463]{Dipen Sahu}
\affiliation{Physical Research Laboratory, Navrangpura, Ahmedabad, Gujarat, PIN-380009}
\email{dsahu@prl.res.in}

\author[0000-0003-4615-602X]{Ankan Das}
\affiliation{Institute of Astronomy Space and Earth Science, P177, CIT Road, Scheme 7m, Kolkata, 700054, West Bengal, India}
\affiliation{Max-Planck-Institute for extraterrestrial Physics, P.O. Box 1312 85741 Garching, Germany}
\email{ankan.das@email.com}

\author[0000-0003-1602-6849]{Prasanta Gorai}
\affiliation{Rosseland Centre for Solar Physics, University of Oslo, PO Box 1029 Blindern, 0315, Oslo, Norway}
\affiliation{Institute of Theoretical Astrophysics, University of Oslo, PO Box 1029 Blindern, 0315, Oslo, Norway}
\email{prasanta.astro@gmail.com }

\author[0000-0002-2887-5859]{Víctor M. Rivilla}
\affiliation{Centro de Astrobiología (CAB), INTA-CSIC, Carretera de Ajalvir, km 4, Torrejón de Ardoz, 28850 Madrid, Spain}
\email{victor.rivilla@cab.inta-csic.es}

\author[0000-0003-4603-7119]{Sheng-Yuan Liu}
\affiliation{Academia Sinica Institute of Astronomy and Astrophysics, 11F of AS/NTU Astronomy-Mathematics Building, No.1, Sec. 4, Roosevelt Rd, Taipei 10617, Taiwan, R.O.C.}
\email{syliu@asiaa.sinica.edu.tw}

\author[0000-0002-2833-0357]{Bhalamurugan Sivaraman}
\affiliation{Physical Research Laboratory, Navrangpura, Ahmedabad, Gujarat, PIN-380009}
\email{bhala.acsi@gmail.com}

\author[0000-0003-1481-7911]{Paola Caselli}
\affiliation{Max-Planck-Institute for extraterrestrial Physics, P.O. Box 1312 85741 Garching, Germany}
\email{caselli@mpe.mpg.de}

\begin{abstract}

The search for dithioformic acid (t-HC(S)SH) in star-forming regions is crucial for understanding interstellar sulfur chemistry and addressing the “missing sulfur” problem. Motivated by a recent claim of t-HC(S)SH detection in NGC 1333 IRAS~4A2, we independently reanalyzed the same ALMA dataset using comprehensive spectral and chemical modeling. We find no credible evidence for t-HC(S)SH: all reported transitions are fully accounted for by known, abundant molecules, with no unblended features unique to t-HC(S)SH. We critically reassess the reported detection, deriving stringent upper limits on the column density ($N_\mathrm{t-HC(S)SH} \leq 4 \times 10^{14}$ cm$^{-2}$) and the fractional abundance ($\leq 1 \times 10^{-10}$ relative to H$_2$). Our astrochemical models place these limits in context, showing the claimed detection likely results from spectral blending and inconsistent modeling assumptions. The non-detection aligns with chemical expectations given the rarity of complex and doubly sulfur-substituted molecules in hot corino environments. Furthermore, our analysis establishes a rigorous framework to guide future searches for sulfur-bearing species and highlights the critical importance of thorough line identification and modeling practices in astrochemistry.

\end{abstract}

\section{ Introduction}

The interstellar medium (ISM) hosts a rich and complex chemistry, yet the full inventory of its molecular constituents remains incomplete. A long-standing challenge is the ``missing sulfur problem'': despite being the tenth most abundant element, only a small fraction of the cosmic sulfur budget is accounted for by known interstellar molecules \citep{Mifsud2021SSRv..217...14M, Jimenez2011A&A...536A..91J, Martin2016A&A...585A.112M, Perdigon2021A&A...647A.162P, Fuente2019A&A...624A.105F}. While dedicated searches have recently expanded the census of S-bearing species with detections of \ch{S2H} \citep{Fuente2017ApJ...851L..49F}, \ch{HCS} \citep{Agundez2018A&A...611L...1A}, \ch{HOCS+} \citep{Sanz-Novo2024ApJ...965..149S}, \ch{HNSO} \citep{Sanz-Novo2024ApJ...965L..26S}, \ch{CH3SCH3} \citep{Sanz-Novo2025ApJ...980L..37S}, \ch{CH3CHS} \citep{Agundez2025A&A...693L..20A},  \ch{CaS} \citep[and tentative presence of KS and KSH in G351.77-mm1,][]{Tasa2025ApJ...993L..42T},   and various sulfur-bearing carbon chains \citep{Cernicharo2021A&A...646L...3C, Cernicharo2021A&A...648L...3C, Cernicharo2021A&A...650L..14C, Cabezas2024A&A...686L...3C}, none of these molecules represent a major sulfur reservoir. For instance, in the cold, dense environment of TMC-1, the total abundance of all detected S-bearing molecules accounts for only 0.15\% of the cosmic sulfur value \citep{Agundez2025A&A...693L..20A}.  Despite S being largely missing in the gas phase in TMC-1, there is a large abundance of new S-bearing molecules detected in this dark cloud 
including the recent detection of c-\ch{C3H2S} \citep{Remijan2025ApJ...982..191R}, and are often referred to as efficient S-bearing molecule factories. Furthermore, Galactic-centre sources such as G+0.693$-$0.027 provide another S-rich environment.  This source hosts multiple S-bearing COMs and thiols, 
including \ch{HC(O)SH} and \ch{C2H5SH} \citep{Rodriguez2021ApJ...912L..11R}, \ch{c-C6H6S} \citep{Araki2026NatAs.tmp...23A}, \ch{CH3SCH3} \citep{Sanz-Novo2025ApJ...980L..37S}, and the metal sulfides NaS and MgS \citep{Rey-Montejo2024ApJ...975..174R}.

The chemical relationship between sulfur and oxygen, which share similar valence electron configurations, provides a useful framework for understanding this problem. Many sulfur-containing molecules have oxygenated counterparts (e.g., \ch{H2CO} vs.\ \ch{H2CS}; \citealt{Agundez2013ChRv..113.8710A, Loison2016MNRAS.456.4101L, Cernicharo2024A&A...682L...4C,roy2025}). Although oxygen is about an order of magnitude more cosmically abundant than sulfur \citep{Fuente2017ApJ...851L..49F}, the abundance ratios of O/S analogs are not straightforward. In some cases, such as the carbon chains C$_2$S and C$_3$S  observed in dark clouds, the sulfur variants are more abundant than their oxygen counterparts, possibly due to sulfur's greater tendency for catenation \citep{Agundez2025A&A...693L..20A} However, for complex organic molecules like \ch{CH3OH} and \ch{C2H5OH}  detected in hot corinos and Galactic-centre clouds, the oxygen-bearing species are typically found to be more abundant \citep{mull16, Sanz-Novo2025ApJ...980L..37S}.

In this context, dithioformic acid (\ch{HC(S)SH}), a thiol\footnote{A thiol is an organosulfur compound characterized by the presence of a (–SH) functional group.} and the fully sulfurated counterpart of formic acid  \citep[\ch{HCOOH},][]{Irvine1990A&A...229L...9I, Jorgensen2018A&A...620A.170J}, represents a key target. The recent detection of its partially sulfurated analog, monothioformic acid (t-\ch{HC(O)SH}), in the Galactic Center cloud G+0.693-0.027 with an abundance of $\sim 10^{-10}$ \citep{Rodriguez2021ApJ...912L..11R},  as well as in the hot core G31.41+0.31 \citep{Garcia2022A&A...658A.150G}, has intensified the search for \ch{HC(S)SH}. Recently, \citet{Manna2024ESC.....8.2401M} claimed the detection of t-\ch{HC(S)SH} toward the hot corino NGC 1333 IRAS 4A2 with a remarkably high abundance of $\sim 2.5 \times 10^{-9}$. This value is comparable to those of the most abundant complex organics in this source, such as glycolaldehyde (\ch{CH2OHCHO}), acetaldehyde (\ch{CH3CHO}), and methyl formate (\ch{HCOOCH3}) \citep{Frediani2025A&A...695A..78F}. The claim is particularly interesting because its oxygenated chemical relative, \ch{HC(O)SH}, whose abundance is expected to be higher than that of \ch{HC(S)SH}, has not been reported in any hot corino \citep{Sanz-Novo2024ApJ...965..149S}.

However, this extraordinary claim warrants careful scrutiny. A close inspection of the analysis by \citet{Manna2024ESC.....8.2401M} reveals several critical inconsistencies that cast doubt on the identification. Observationally, all selected t-\ch{HC(S)SH} transitions share nearly identical upper-state energies see Table~\ref{tab:HC(S)SH}), which precludes a reliable determination of the rotational temperature and introduces significant degeneracy into the column density calculation. Furthermore, the analysis does not sufficiently account for severe spectral blending from abundant, known hot corino species. On the theoretical side, the supporting chemical model is not benchmarked against other observed molecules and, most crucially, compares a predicted \textit{ice-phase} abundance with \textit{gas-phase} observational data—a fundamentally flawed comparison. 

Given these profound concerns, we undertook an independent and rigorous reanalysis of the same archival ALMA data presented in \citet{Sahu2019ApJ...872..196S} and \cite{Su2019ApJ...885...98S}. Our work provides a comprehensive re-evaluation through a transition-by-transition analysis to robustly identify all spectral features near the claimed \ch{HC(S)SH} frequencies. We establish a stringent upper limit for the column density of t-\ch{HC(S)SH} and benchmark it against chemically related species. Finally, we apply an advanced gas-grain chemical model, validated against known molecular abundances, to predict the expected abundance of t-\ch{HC(S)SH} under hot corino conditions.

Our analysis demonstrates that t-\ch{HC(S)SH} is not detected in IRAS 4A2. The features previously attributed to it are fully explained by line blending for other common chemical species. Our derived upper limit and chemical models are mutually consistent and suggest that the  actual abundance is at least an order of magnitude lower than the prior claim. This work refutes a significant detection claim, constrains the abundance of a key sulfur species, and underscores the critical importance of rigorous, multi-faceted analysis in spectrally dense star-forming regions.

The paper is organized as follows: Section 2 describes the data and our analysis methods. Section 3 presents the results of our spectral line identification and upper limit calculations. Section 4 details our astrochemical modeling. Section 5 discusses the implications of our findings, and Section 6 provides our conclusions.

\section{Data analysis } \label{sec:data}

\subsection{ALMA observations}
The observations of IRAS 4A were conducted using the Atacama Large Millimeter/Submillimeter Array (ALMA) under project code 2015.1.00147.S (P.I. Yu-Nung Su). Data collection involved three Execution Blocks (EBs), conducted on July 23 and 24, and December 14, 2016, using the ALMA Band 7 receiver with array configurations C40-3 and C40-5.

The phase center for these observations was set at a right ascension (RA) of $\rm{03^h29^m10\fs50}$ and a declination (Dec) of $+31\degr13\arcmin31\farcs5$ (J2000). The total on-source integration time was approximately 84 minutes. A total of seven spectral windows were deployed, including a broadband window centered at a frequency of 350.714\,GHz, which featured a bandwidth of 1.875\,GHz and a spectral channel width of 976\,kHz. The application of standard online Hanning smoothing yielded a velocity resolution of approximately 0.84\,\kms. This study focuses primarily on data from the aforementioned broadband window; additional details on the full observational setup are discussed in \citet{Sahu2019ApJ...872..196S, Su2019ApJ...885...98S}.

The initial calibration was performed by ALMA using the pipeline within the Common Astronomy Software Application (CASA) package, version 4.7 \citep{McMullin2007ASPC..376..127M}. To address the goals of this work, we reprocessed the data to generate two distinct sets of spectral cubes corresponding to different array configuration: a high-resolution set $\sim 0.20\arcsec$ and a coarser-resolution set $\sim 0.55\arcsec$.  Images were formed using Briggs weighting with a robustness parameter of 0.5. Continuum subtraction was performed using \texttt{statcont} \citep{Sanchez2018A&A...609A.101S}. The resulting synthesized beam size for the high-resolution images is $0\farcs24 \times 0\farcs17$ (P.A. = $-4.3\degr$), while the coarser-resolution images have a beam size of $0\farcs66 \times 0\farcs47$ (P.A. = $-11.8\degr$). 

The resulting root mean square (rms) noise levels for the high-resolution and coarser-resolution datasets are 3 and 12\,mJy\,beam$^{-1}$ for the continuum, 
and $4$--$8\,\mathrm{mJy\,beam^{-1}}$ ($ 1\, \mathrm{mJy\,beam^{-1} \equiv 0.24~\mathrm{K}}$) and $4$--$12\,\mathrm{mJy\,beam^{-1}}$ ($\mathrm{ 1 \, mJy\,beam^{-1} \equiv 0.032~K}$) for the spectral lines, respectively.
The noise levels in both datasets are primarily limited by the dynamic range due to bright emission features in the source.

Complex organic molecules (COMs) in this source originate primarily from the compact hot corino region, spanning approximately 70~au ($\sim 0\farcs24$), though recent studies suggest an even more compact distribution \citep{Frediani2025A&A...695A..78F}. Consequently, using the coarser resolution data is suboptimal for reporting new molecular detections, as it suffers from significant beam dilution, rendering spectral features weaker than in the $0\farcs2$ observation. It is unclear why \citet{Manna2024ESC.....8.2401M} relied exclusively on coarser-resolution data while omitting higher-resolution observations, especially since they reported that the \ch{t-HC(S)SH} emission was unresolved. Therefore, we prioritize the high-resolution data for our spectral line analysis. We also process the coarser-resolution datasets to ensure internal consistency and to provide a direct benchmark against the results reported by \citet{Manna2024ESC.....8.2401M}. To extract the spectra uniformly, we first performed a 2D Gaussian fit on the continuum emission using \texttt{imfit}, determining a central position of $\alpha_{J2000} = 03^h29^m10\fs4325$, $\delta_{J2000} = +31\degr13\arcmin32\farcs0585$. Based on this position, spectra were extracted from circular regions with radii of $0\farcs1$ and $0\farcs3$  for the high- and low-resolution cubes, respectively, corresponding to their average beam sizes. 

In addition, continuum maps are used to estimate the average column density ($N_{H_2}$) for these regions.  We assumed a dust opacity formulation of $\kappa_\nu = 0.6(\nu/245\text{GHz})^{1.5}$ and a standard gas-to-dust mass ratio of 100, consistent with previous studies of this source \citep{Sahu2019ApJ...872..196S, Su2019ApJ...885...98S} Additionally, a representative dust temperature of 60~K was adopted for the core, which is similar to the peak brightness temperature of IRAS 4A1. This assumption is justified because the beam-averaged peak continuum brightness temperature observed specifically toward 4A2 is 42~K \citep{Su2019ApJ...885...98S}. Since the dust emission toward IRAS 4A2 is predominantly optically thin, this observed value serves as a lower limit, implying that the actual dust temperature is likely even higher.
The estimated column densities for the high and coarser resolution data are $2.5 \times 10^{25}$\,cm$^{-2}$ and $4.9 \times 10^{24}$\,cm$^{-2}$, respectively. These values are used to estimate the molecular abundances, assuming the emission originates from the same region without additional dilution factors.

\subsection{HC(S)SH : Laboratory spectroscopic information}
HC(S)SH isomers are classified as near-prolate asymmetric rotors, with the asymmetry parameters given by \( \kappa_{\text{trans}} = -0.9901 \) and \( \kappa_{\text{cis}} = -0.9682 \) \citep{Prudenzano2018A&A...612A..56P}. Both isomers exhibit \( C_s \) symmetry and possess a planar structure. The \( a \) components of the dipole moment have been experimentally determined by \citet{Bak1979JMoSp..75..134B}, resulting in \( \mu_a(\text{trans}) = 1.53 \, D \) and \( \mu_a(\text{cis}) = 2.10 \, D \). The \( b \) components were  computed  by \citet{Prudenzano2018A&A...612A..56P} and reported to be 0.19 D for the trans  isomer and 1.64 D for the cis. Although \citet{Bak1979JMoSp..75..134B} conducted the first laboratory measurement, their measurements were limited to the centimeter (cm) regime, specifically between 18 and 40 GHz. 

  \citet{Prudenzano2018A&A...612A..56P} subsequently measured 204 new line frequencies for the trans  isomer and 139 for the cis isomer,  with an accuracy within 5–50 kHz in the mm and submm ranges. The accuracy was at least an order higher than the previous study \citep{Bak1979JMoSp..75..134B}. Consequently, the measured spectra are now available in bands such as ALMA 6 and 7. The new measurements cover a wide range of quantum numbers \( J \) and \( K_a \).

These updated measurements are cataloged in the Cologne Database for Molecular Spectroscopy \citep[CDMS;][]{Muller2005JMoSt.742..215M}, enabling robust astronomical searches across centimeter, millimeter, and submillimeter wavelengths. The expected peak emission frequency strongly depends on the source temperature. For lower temperatures around 10 K, typical of cold environments like prestellar cores, the trans isomer exhibits a peak around 45 GHz. In warmer regions with kinetic temperatures \( T_{\text{kin}} \sim 100 \, K \), this peak shifts to higher frequencies, approximately 200 GHz. Due to its larger \( b \) component of the dipole moment, the cis isomer displays a different intensity distribution; in this case, the calculated peak occurs at approximately 280 GHz in cold sources and at around 780 GHz in warmer regions. However, the strongest lines above 300 GHz—specifically b-type transitions with low \( E_u/k \)—exhibit uncertainties greater than 1 MHz. These significant errors arise from the limited number of measurable b-type transitions, which are constrained by the overall lower intensity of the cis lines compared to those of the trans isomer.

\subsection{Line Identification and Analysis Approach}

We performed line identification and modeling using the CASSIS software package \citep{Vastel2015sf2a.conf..313V} under the assumption of Local Thermodynamic Equilibrium (LTE), a standard approach for hot corino studies \citep[e.g.,][]{Jorgensen2018A&A...620A.170J}.  Because the compact hot corino emission may be affected by beam dilution, a proper source size should be considered for LTE modeling. We kept the source size as a free parameter in the CASSIS models and individually checked whether the fitted output source size was smaller than the beam size. Unless the source size is smaller than the beam size, it does not affect the resultant synthetic spectra. The source sizes for the best-fit cases are listed in the tables of Appendix C. Our core principle is that a reliable identification requires the synthetic spectrum of a species to simultaneously match observed features and not predict strong, unobserved lines elsewhere in the band. Given the high line density in IRAS~4A2, which can lead to significant spectral confusion, we adopted a multi-step validation strategy to ensure the robustness of our analysis before addressing the reported t-\ch{HC(S)SH} claim.

First, we established a baseline chemical inventory by prioritizing the identification of known, abundant hot corino species. We leveraged the close spectral similarity between IRAS~4A2 \citep{Sahu2019ApJ...872..196S, Sahu2020ApJ...899...65S} and the well-studied hot corino IRAS~16293-2422, using the ALMA-PILS survey \citep{Jorgensen2018A&A...620A.170J} results as a 
guide  to account for the most likely overlapping emission from abundant hot‑corino species.

Second, to validate our LTE modeling approach for this specific source, we derived column densities for several well-detected species and compared them with previously published values for IRAS~4A2 that were obtained via the rotational diagram method \citep{Sahu2019ApJ...872..196S}. Our results were consistent, confirming that our LTE modeling accurately reproduces established abundances for relatively unblended lines. 
 For abundant, optically thick species such as \ch{CH3OH}, we derived the column density from LTE/MCMC fits to optically thin transitions of their less abundant isotopologues (e.g.\ \ch{^{13}CH3OH} and \ch{CH3^{18}OH}), and then scaled by an assumed typical interstellar isotopic ratios (e.g.\ ${}^{12}\mathrm{C}/{}^{13}\mathrm{C}$, ${}^{16}\mathrm{O}/{}^{18}\mathrm{O}$).

This rigorous and validated framework allowed us to construct a reliable chemical inventory and model the contribution of all confirmed species.  Only after this comprehensive accounting did we proceed to search for any remaining spectral evidence of t-\ch{HC(S)SH}.
 In this connection, it is relevant to mention that \citet{Synder2005ApJ...619..914S} in their refutation of interstellar glycine emphasized the necessity of accurate rest frequencies, careful consideration of line confusion and blending, identification of connecting transitions, and consistent intensity of transitions of equal strength, unless non-thermal excitation processes apply. 
They emphasized the importance of evaluating the measured column densities of proposed molecules to ensure consistency with abundances of well-characterized species. Recently, \citet{Xue2019ApJ...882..118X} quantitatively advanced this framework by rigorously assessing line contamination and line intensity matching in line-crowded  spectra, particularly for the identification of new molecules. We not only considered all of those criteria, but further performed the chemical modeling to understand the chemical origin and expected abundance.  
All details are presented in the following  Section~\ref{sec:HC(S)SH}.


\section{Results}
In this section, we present the results of our spectral analysis. We first re-examine the reported transitions of t-\ch{HC(S)SH}, demonstrating that they are attributable to line blending. We then derive an upper limit for the column density of t-\ch{HC(S)SH} and provide crucial chemical context by searching for related species \ch{HC(O)SH}. 

\subsection{Re-examining the Claimed Detection of t-HC(S)SH} \label{sec:HC(S)SH}
The observed ALMA spectral window centered at 350.714\,GHz contains a dense forest of molecular lines \citep{Sahu2019ApJ...872..196S}. Our analysis focuses on the full broadband spectral window: 349.73--351.606\,GHz range (1.876\,GHz bandwidth),  excluding only a single bad channel (0.97\,MHz) at the band end. Within this band, we identified five transitions of t-\ch{HC(S)SH}, which are listed in Table~\ref{tab:HC(S)SH} and were previously reported by \citet{Manna2024ESC.....8.2401M}. 

As described in Sec.~\ref{sec:data}, two distinct spectral datasets were employed for this analysis. We primarily focus on the high angular resolution ($\sim 0.2\arcsec$) spectra to conduct an exhaustive inventory of the molecular species in the vicinity of the t-\ch{HC(S)SH} transitions (Fig.~\ref{fig:spectra1}). Subsequently, we applied the same analytical techniques to the lower angular resolution ($\sim 0.55\arcsec$) spectra, presenting only the dominant species that are confirmed to blend with the reported t-\ch{HC(S)SH} features (Fig.~\ref{fig:spectra2}).
 

\begin{table*}
\caption{Transitions of t-\ch{HC(S)SH} within the observed spectral range.}
\label{tab:HC(S)SH}
\centering
\begin{tabular}{cccccc}
\hline
\hline
No. & Species & Quantum Numbers & Frequency (GHz) & $E_u$ (K) & $\log_{10}(A_{ij} / \text{s}^{-1})$ \\
\hline
1 & \ch{t-HC(S)SH} & 54(1,54)-53(1,53) & 350.3516 & 466.4 & -3.3 \\
2 &\ch{t-HC(S)SH} & 52(3,49)-51(3,48) & 350.4011 & 463.0 & -3.3 \\
3 &\ch{t-HC(S)SH} & 54(0,54)-53(0,53) & 350.4996 & 466.4 & -3.3 \\
4 & \ch{t-HC(S)SH} & 52(2,50)-51(2,49) & 350.8495 & 454.6 & -3.3 \\
5 & \ch{t-HC(S)SH} & 53(1,52)-52(1,51) & 351.5882 & 462.9 & -3.3 \\
\hline
\end{tabular}
\end{table*}

A critical issue with these  five candidate lines is immediately apparent from their spectroscopic properties (Table~\ref{tab:HC(S)SH}). They all possess nearly identical upper-state energies ($E_u \approx 460$\,K) and  Einstein coefficients (log $A_{ij} \approx -3.2$). This degeneracy makes a reliable determination of the rotational temperature via a rotational diagram impossible. It also implies that under LTE conditions,  any unblended emission from these lines should have a similar intensity (see Fig.~\ref{fig:LTE_upperlimit}). As we explained below, the observed spectral features do not meet this condition, and their origins may be traced to other, well-known hot corino molecules.

To properly assess potential blending, we first established a comprehensive chemical inventory of the source. The line-rich nature of the spectrum is illustrated in Figure~\ref{fig:spectra1}, where we show the contributions of the most significant molecules we identified. Our inventory includes simple species and their isotopologues, such as formic acid (\ch{HCOOH}), isocyanic acid (\ch{HNCO} and \ch{HN^{13}CO}), and methanol (\ch{CH3OH} and \ch{^{13}CH3OH}), as well as numerous complex organic molecules like methyl formate (\ch{CH3OCHO}), acetaldehyde (\ch{CH3CHO}) glycolaldehyde (\ch{CH2OHCHO}), ethanol (\ch{C2H5OH}), ethyl cyanide (\ch{C2H5CN}), and ethylene glycol (\ch{aGg-(CH2OH)2}).  Additionally, we search for previously reported species like formic acid (\ch{HCOOH})
 whose strong transitions fall outside our band.

To ensure that our analysis was exhaustive, we also systematically searched for a wide range of other potential  species \citep[often referred to as spectral `weeds' in line-crowded regions; e.g.,`weeds'][]{Fortman2010ApJ...725L..11F}  but did not find conclusive evidence for their presence. A complete list of these searched-for but unconfirmed species is provided in Appendix~\ref{app:searched_molecules}. Having established this rich molecular background, we proceeded to analyze each of the  five t-\ch{HC(S)SH} candidate lines individually.

\subsubsection{Uncovering the Origin of Emission near t-HC(S)SH Frequencies}
To identify the true origin of the emission features, we conducted a comprehensive search for all cataloged molecular transitions within $\pm25$\,\kms of each t-\ch{HC(S)SH} line, which include well known molecules often detected toward hot corinos. Our analysis reveals that the targeted emission
spectra are dominated by a handful of abundant complex organic molecules, including methyl formate (\ch{CH3OCHO}), glycolaldehyde (\ch{CH2OHCHO}), acetaldehyde (\ch{CH3CHO}), and ethylene glycol (\ch{g'Ga-(CH2OH)2});  detailed LTE synthetic spectra for these individual species are presented in Appendix D. Figure~\ref{fig:spectra1} provides a comprehensive overview of the total model, displaying all identified species alongside the major blending species. In contrast,  Figure~\ref{fig:spectra2} displays the data at a coarser angular resolution, isolating only the major species responsible for the target line blending.)

A detailed breakdown of each transition reveals the extent of the spectral contamination:
\begin{figure*}[h!]
 \centering
 \includegraphics[width=1.0\linewidth]{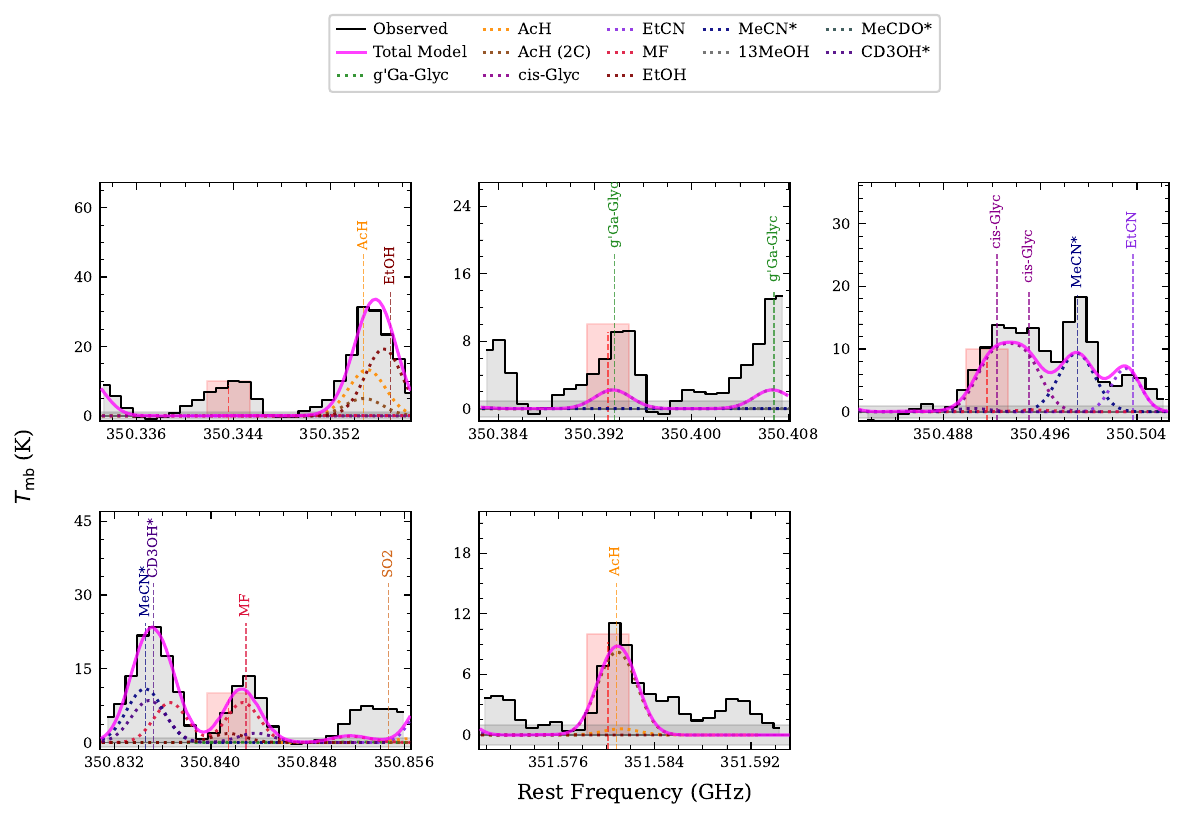} 
 \caption{Comprehensive spectral analysis of the regions surrounding the four t-\ch{HC(S)SH} transition frequencies. The observed spectrum is shown in black. The combined synthetic model, accounting for all identified molecules, is shown in magenta. Individual contributions from the most significant species are overplotted in various colors, showing that known molecules mostly account for the observed emission features. The shaded red region show bar (height 10~K) around the t-\ch{HC(S)SH} transition with a vlsr=6.9 \kms with a total width of 3~\kms ($\pm 1.5$ \kms).  The dimgray horizontal shadow shows $\pm 1\sigma$ (rms) around the zero baseline.
 \newline
Note. ---  The following abbreviations are used throughout the figures: MF (CH$_3$OCHO), cis-Glyc (CH$_2$OHCHO), g'Ga-Glyc (g'Ga--(CH$_2$OH)$_2$), AcH (CH$_3$CHO), EtOH (C$_2$H$_5$OH), EtCN (C$_2$H$_5$CN), MeCN* (CH$_3$CN $v_8=1$), 13MeOH ($^{13}$CH$_3$OH), CD3OH* (CD$_3$OH $v_t=1$), and MeCDO* (CH$_3$CDO $v_t=0,1$).}

 \label{fig:spectra1}
\end{figure*}

\begin{figure*}[h!]
 \centering
 \includegraphics[width=1.0\linewidth]{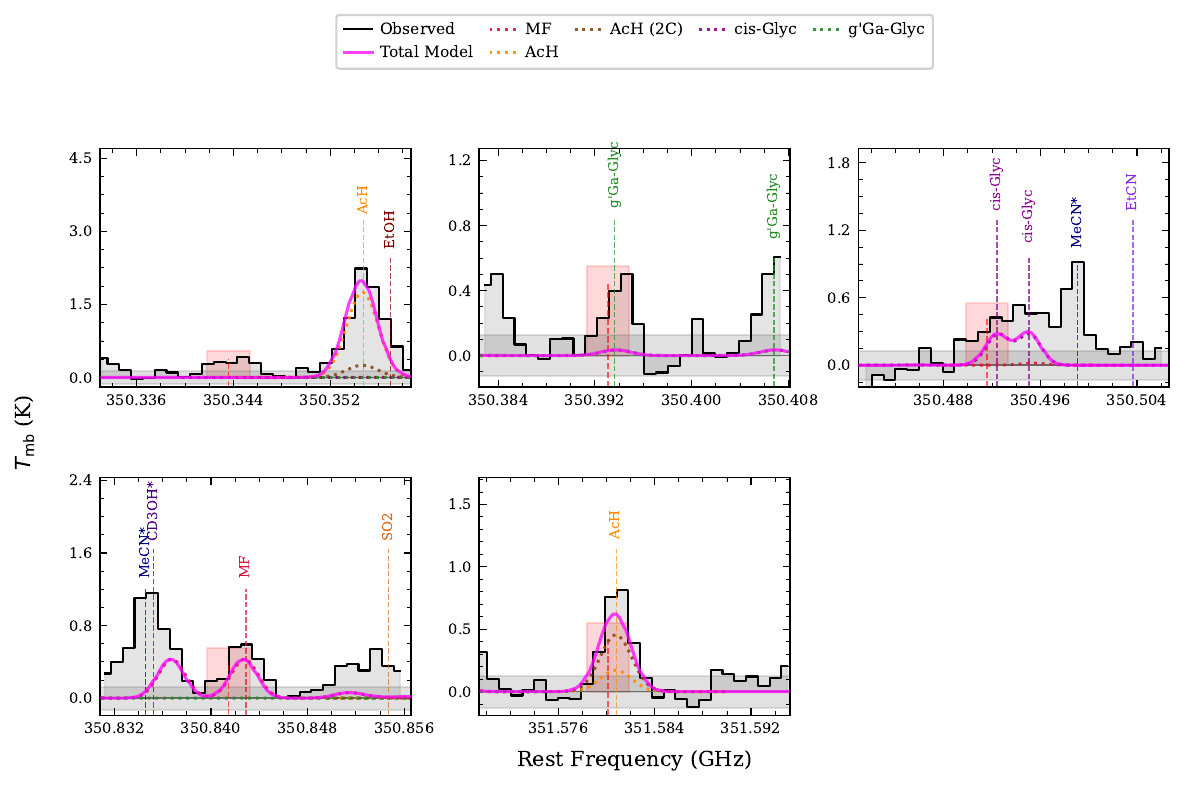} 
 \caption{Similar to Figure~\ref{fig:spectra1}, but showing the data at a coarser angular resolution ($0\farcs55$), which closely matches the resolution analyzed in the previous report \citep{Manna2024ESC.....8.2401M}. Note that the `Total Model' shown here comprises only the selected major blending species listed in the legend.}
 \label{fig:spectra2}
\end{figure*}

\paragraph{Transition 1 (350.3516 GHz):} The observed feature at this frequency is comparatively weak and cannot be fully explained by any single, confirmed species. While vibrationally excited methyl cyanide (\ch{CH3CN}, $v_8=1$) provides a minor contribution  (see Appendix D), the peak of the emission remains unidentified. However, this feature cannot originate from t-\ch{HC(S)SH}. As established in Section~\ref{sec:HC(S)SH}, all five candidate transitions of t-\ch{HC(S)SH} possess nearly identical spectroscopic parameters, which means they produce lines of comparable intensity under LTE. As transitions 3,4 and 5  clearly indicate little or no contribution of  t-\ch{HC(S)SH}, the unexplained emission feature around this transition is not arising from t-\ch{HC(S)SH}.

\paragraph{Transition 2 (350.4011 GHz):}
The emission feature at this frequency is significantly contaminated by a known transition of ethylene glycol (\ch{g'Ga-(CH2OH)2}). While the remaining emission peak is not definitively identified, it cannot be attributed to t-\ch{HC(S)SH}. Following the same logic as for  transition 1, the intensity of this residual feature is inconsistent with the expected uniform line strength of the four t-\ch{HC(S)SH} transitions, especially when compared to the features at the frequencies of transitions 3 and 4. We note that while a transition of disulfur monoxide (\ch{S2O}) could potentially explain the residual peak, \ch{S2O} has not been confirmed in the ISM, and we find no other corroborating transitions in our data to support such a claim. Therefore, we do not consider it a plausible identification and have not included it in our combined synthetic/model spectra.

\paragraph{Transition 3 (350.4996 GHz):}
The emission feature at this frequency is unequivocally dominated by the well-known hot corino molecule glycolaldehyde (\ch{CH2OHCHO}). We confirmed this identification with a comprehensive analysis, performing an MCMC fit on all 22 glycolaldehyde transitions present within our spectral window. 
Notably, the observed feature is not a single, narrow line but a broad profile created by the blending of multiple strong glycolaldehyde transitions. The synthetic spectrum generated from our robust fit for glycolaldehyde alone perfectly reproduces the observed emission, leaving almost no residual signal that could be attributed to t-\ch{HC(S)SH}. This serves as a definitive refutation of this line as a candidate for t-\ch{HC(S)SH}.

\paragraph{Transition 4 (350.8495 GHz):} This feature is entirely explained by a strong transition of vibrationally excited methyl formate (\ch{CH3OCHO}, $v=1$). Our modeling of multiple methyl formate transitions across the band produces a good match with the observed spectra. This clearly confirms \ch{CH3OCHO} as the sole origin of the observed feature around this transition.

\paragraph{Transition 5 (351.5882 GHz):} Similar to transitions 3 and 4, the spectral emission around this frequency is primarily dominated by strong \ch{CH3CHO} emission. As shown in the tables in the appendix, a strong transition exists (log A$_{ij} \sim -3.9$) with E$_u$=39.8\,K. While a single high excitation temperature ($> 100$ K) cannot reproduce the emission feature in LTE modeling, the observed profile is closely explained by a cold component of \ch{CH3CHO} assuming a low T$_{\rm{ex}}$ of 30\,K.

Since the features at the frequencies of transitions 3, 4 and 5 are unequivocally produced by glycolaldehyde, methyl formate,  and acetaldehyde respectively, and because all five t-\ch{HC(S)SH} transitions should have similar intensities, we conclude that the reported presence of **t-\ch{HC(S)SH}**  can not be claimed as detection.

\subsection{Upper Limit for t-HC(S)SH Abundance}
Based on the non-detection, we derive a stringent upper limit on the column density of t-\ch{HC(S)SH}, assuming an uncertainty of $\sim$20\% in the synthetic LTE spectra. 

First, we attempted to reproduce the spectra using the parameters reported by \citet{Manna2024ESC.....8.2401M}: $N_{\text{t-HC(S)SH}} = 2.5 \times 10^{15}$\,\pcms, $T_\mathrm{ex} = 255$\,K, and FWHM = 3.2\,\kms. Although the previous authors adopted a central velocity of $V_\mathrm{LSR} = 6.3$\,\kms, we utilized $V_\mathrm{LSR} = 6.9$\,\kms, consistent with the value reported for this source by \citet{Sahu2019ApJ...872..196S}. We found that a column density of $N_{\text{t-HC(S)SH}} \approx 2.0 \times 10^{15}$\,\pcms\ is sufficient to match the peak intensity levels of the observed features, confirming that our spectral synthesis is similar with the previous report.

\begin{figure*}[h!]
 \centering
 \includegraphics[width=1.0\linewidth]{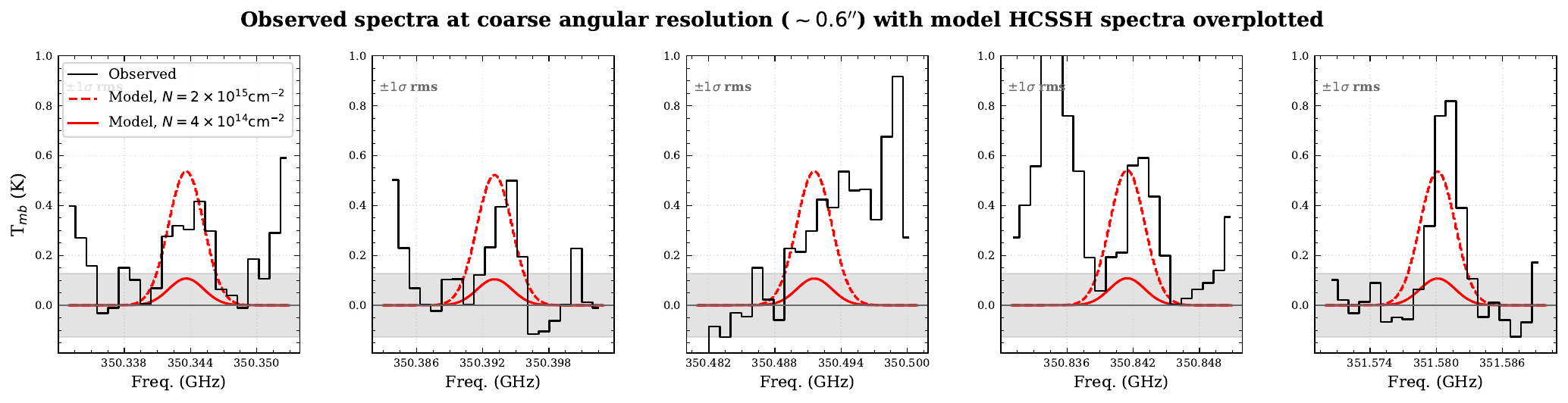} 
 \caption{Model t-\ch{HC(S)SH} overplotted with observed spectra. Synthetic LTE spectra are overplotted with column density values close to value reported by previous report and possible upper limit based on our study. The dimgray horizontal shadow shows $\pm 1\sigma$ (rms) around the zero baseline. }
 \label{fig:LTE_upperlimit}
\end{figure*}

However, when accounting for the spectral uncertainties  of about 20\% and the absence of clean, unblended features, the data constrain the maximum possible column density to $N_{\text{t-HC(S)SH}} \leq 4 \times 10^{14}$\,\pcms\ (see Fig.~\ref{fig:LTE_upperlimit}). 
 Although we derived H$_2$ column densities for both the high-resolution ($0\farcs20$) and coarser-resolution ($0\farcs55$) datasets (see Section 2.1), we explicitly note the value derived from the $0\farcs55$ data to calculate the upper limit. This choice ensures a direct comparison with the fractional abundance reported by \citet{Manna2024ESC.....8.2401M}, who relied exclusively on data at this coarser resolution.
For the $0\farcs55$ resolution observations, adopting an \ch{H2} column density of $N_{\text{H}_2} = 4.9 \times 10^{24}$\,\pcms, the corresponding fractional abundance is:
\begin{equation}
    X_{\text{t-HC(S)SH}} \leq 1 \times 10^{-10}
\end{equation}
This upper limit is nearly an order of magnitude lower than the abundance of $2.5 \times 10^{-9}$ claimed by \citet{Manna2024ESC.....8.2401M}. We emphasize that this upper limit calculation does not imply even a tentative detection. These results are also consistent with the expectations from our chemical models (see Section 4)..

\subsection{Constraints from a Chemically Related Species: HC(O)SH} \label{sec:hcosh}

To provide further chemical context for the non-detection of t-\ch{HC(S)SH}, we searched for its less sulfurated analog, monothioformic acid (\ch{HC(O)SH}), which is expected to be more abundant.   While establishing the chemical context of oxygenated analogs, we also searched for the fully oxygenated species, formic acid (HCOOH). However, no strong transitions of HCOOH are present within our observed spectral range. Nevertheless, HCOOH has been previously detected toward this source by \citet{Lopez2017A&A...606A.121L} with an observed abundance of $(0.6\text{--}2.9) \times 10^{-9}$. This previous observational constraint closely matches with our astrochemical modeling, which reproduces a peak fractional abundance of $2.3 \times 10^{-9}$ for HCOOH (see Section~\ref{subsec:modeling_results}). The \emph{expected} higher abundance of  HC(O)SH is primarily because oxygen is more cosmically abundant than sulfur and because analogous  O/S pairs (e.g.\ HCOOH vs.\ HCSSH) are usually taken to follow that elemental trend, rather than on the basis of a detailed comparison of formation kinetics (also see Sec.~\ref{subsec:modeling_results}). We searched for transitions of both the cis and trans  isomers of \ch{HC(S)SH}.

\begin{figure*}[h!]
 \centering
 \includegraphics[width=0.9\linewidth]{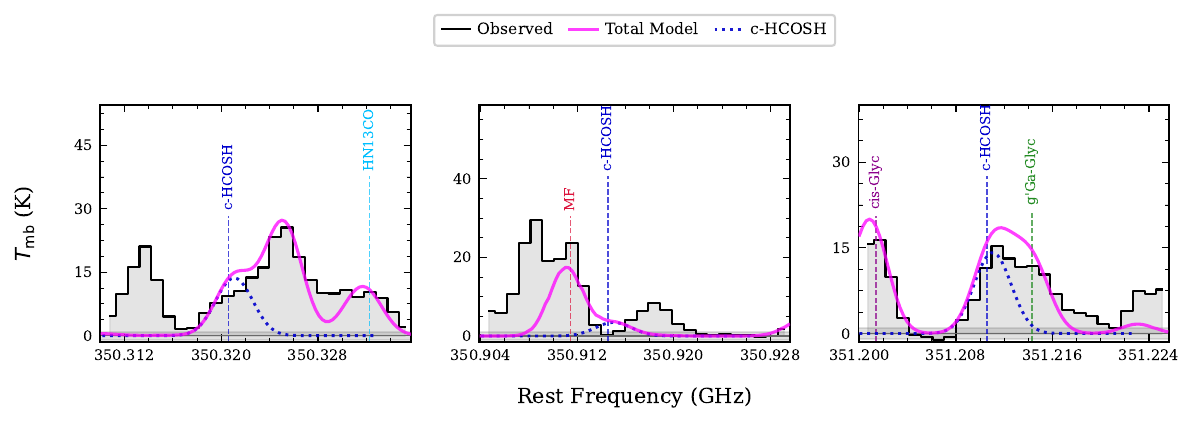}
 \caption{Observed spectra (black) around three transitions of c-\ch{HC(O)SH}. The best-fit synthetic model is overplotted in blue, indicating a tentative detection.}
 \label{fig:cHCOSH}
\end{figure*}

\begin{figure*}[h!]
 \centering
 \includegraphics[width=0.9\linewidth]{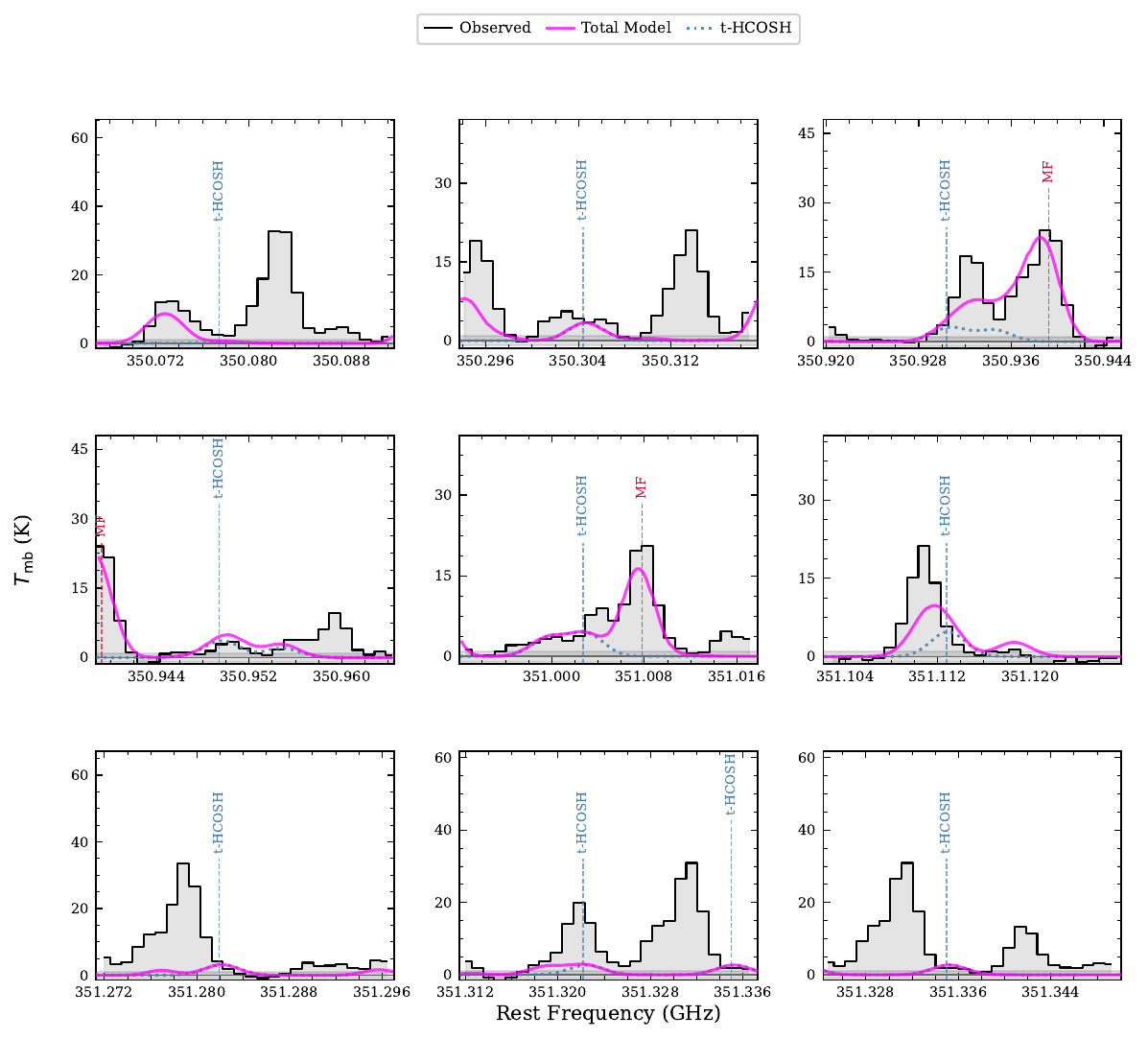}
 \caption{Observed spectra (black) around several transitions of t-\ch{HC(O)SH}. The best-fit synthetic model (blue) suggests a tentative detection at a low abundance level.}
 \label{fig:tHCOSH}
\end{figure*}

For c-\ch{HC(O)SH}, we identified two partially blended transitions (Table~\ref{tab:cHC(O)SH} and Figure~\ref{fig:cHCOSH}), which are consistent with a column density of $N = 1.2 \times 10^{16}$\,\pcms. For t-\ch{HC(O)SH}, we found several weak and partially blended features (Table~\ref{tab:tHC(O)SH} and Figure~\ref{fig:tHCOSH}), corresponding to a column density of $N = 7 \times 10^{15}$\,\pcms. The best-fit parameters from our MCMC analysis are summarized in Table~\ref{app:column_densities}.  Though the spectral features may not imply a confirmed detection,  but they are good enough for upper limit estimation. 

These column densities translate to fractional abundances of $X_{\text{c-HC(O)SH}} \approx 3.2 \times 10^{-10}$ and $X_{\text{t-HC(O)SH}} \approx 1.2 \times 10^{-10}$. These values are consistent with the abundance of $\sim 1 \times 10^{-10}$ found for t-\ch{HC(O)SH} in the Galactic Center cloud G+0.693–0.027 \citep{Rodriguez2021ApJ...912L..11R}. The fact that \ch{HC(O)SH} is present at this low abundance level provides strong chemical evidence against the claim of a much higher abundance for the doubly sulfur species t-\ch{HC(S)SH}.

\section{Astrochemical Modeling}
Earlier astrochemical frameworks have demonstrated the intricate behavior of sulfur-containing compounds, attributed to their diverse and dynamic chemical interactions \citep{Wakelam2004, vida17, Gorai2017, lass19,char97}. The reaction networks typically include uncertain gas-phase and grain-surface processes, posing difficulties for precise modeling. Moreover, the depletion of sulfur and its reallocation across different reservoirs add further complexity to abundance estimation. These intricacies often result in notable discrepancies between theoretical models and observational data.
To theoretically constrain the abundance of \ch{HC(S)SH} and its chemical analogs, we employ the Chemical Model for Molecular Cloud (CMMC) \citep{das15,das19,das21,das25,gora20}. We developed a physical model representative of a hot corino environment and updated our chemical network with the most relevant formation and destruction pathways.

\subsection{Modeling Methodology}

\subsubsection{Physical Setup}

The chemical evolution of \ch{HC(S)SH} and its related species was studied using a physical model representative of a generic hot corino, which is applicable to NGC 1333 IRAS 4A2. 
We consider three distinct evolutionary stages in our chemical modeling. The first is a prestellar phase characterized by a gas density of $2\times10^{4}\,\mathrm{cm^{-3}}$, equal gas and dust temperatures of 8~K, and a visual extinction of $A_V=10$. The second stage corresponds to a free-fall collapse accompanied by a warm-up phase, during which the density increases from $2.0\times10^{4}$ to $10^{7}\,\mathrm{cm^{-3}}$, the temperature rises from 8 to 300~K, and the visual extinction increases with density. The final stage is a post–warm-up phase, in which the density, temperature, and visual extinction are held fixed at the values attained at the end of the collapse and warm-up phase.

This overall physical framework is broadly consistent with the model explored by \citet{Sahu2020ApJ...899...65S}. However, our treatment of the density and temperature evolution during the collapse and warm-up phases is slightly modified following a physical prescription and also to account for two sets of observations we present here for which the column density differs by close to an order of magnitude. While \citet{Sahu2020ApJ...899...65S} assume linear increases in both density (up to $10^8$ cm$^{-3}$) and temperature (up to 300 K) over a collapse timescale of $\sim2.1\times10^{5}$~yr, we instead adopt a density evolution up to a density of $10^7$ cm$^{-3}$ following a free-fall collapse prescription \citep{brown88}, with the temperature increase explicitly coupled to the rise in density.
The visual extinction is directly linked to the evolving density according to
\begin{equation}
A_V = A_{V_0}\left(\frac{n_{\mathrm{H}}}{n_{\mathrm{H}_0}}\right)^{2/3},
\end{equation}
where $n_{\mathrm{H}_0}=2\times10^{4}\,\mathrm{cm^{-3}}$ and $A_{V_0}\simeq10$ correspond to the initial conditions at the onset of collapse \citep{garr11}.

Following the completion of the warm-up phase, the chemical evolution is continued for an additional $10^{5}$~yr at fixed density, temperature, and visual extinction in order to capture post–warm-up chemistry. A standard cosmic-ray ionization rate \citep{herb09} of $\zeta = 1.3\times10^{-17}\,\mathrm{s^{-1}}$ is adopted throughout all stages of the simulation. The initial elemental abundances at the prestellar stage are taken from \citet{Esplugues2013} and are listed in Table~\ref{tab:init}.

\subsubsection{Chemical Network}
Our ice phase chemical network is built upon the framework of \citet{ruau16}, whereas the gas phase network is adopted from \cite{mill22}. Additionally, it includes H-abstraction reactions from \citet{bell14} and \citet{garr17} and has been updated with reactions from our recent works \citep{mond21,mond23,sil21,sriv22,rama24}. For this study, we have incorporated a comprehensive set of formation and destruction pathways for \ch{HC(S)SH} and \ch{HC(O)SH}. The complete list of added reactions is provided in Appendix~\ref{app:reactions} (see Table \ref{table:reactions}).

\paragraph{Gas-phase Reactions}

In addition to our existing network, we incorporate the formation and destruction processes of HC(S)SH and some related species, as listed in Table \ref{table:reactions}. The gas-phase formation and destruction of HC(S)SH and HC(O)SH in our network are based on the available reactions for HCOOH found in the UMIST database for Astrochemistry \citep{mill22}. \citet{lass19} proposed that HC(S)SH may form in the gas phase through the neutral-neutral reaction between HS and H$_2$CS. A comparable reaction is available in the UMIST database for the formation of gas-phase HCOOH, which occurs through the reaction between H$_2$CO and OH, with $\alpha = 2 \times 10^{-13}$, $\beta = 0$, and $\gamma = 0$. Additionally, a faster reaction pathway is considered in the UMIST network, which produces HCO and H$_2$O, with rate constants of $\alpha = 7.73 \times 10^{-12}$, $\beta = -1.03$, and $\gamma = 0$. Here, we include both channels (gas-phase reaction numbers 1 and 3). To determine the upper limit for the formation of HC(S)SH, we also evaluate a case with $\alpha = 10^{-10}$ for gas-phase reaction number 1. Similar rate constants and the upper limit are considered for the formation of HC(O)SH by the gas-phase reaction between  H$_2$CS and OH (gas-phase reaction numbers 2 and 4).

Another gas phase reaction was proposed by \citet{lass19} for the formation of HC(S)SH by the dissociative recombination of $\rm{HC(S)SH_2}^+$ (gas phase reaction number 7). Following the rate constants available for the dissociative recombination reaction of $\rm{HCOOH_2}^+$ and its other channels, we consider similar rate constants here. Similar dissociative recombination reaction channels are considered for $\rm{HC(O)SH_2}^+$, which yields HC(O)SH (gas phase reaction number 9).

For destruction mechanisms, we included: (a) reactions with major interstellar ions ($\rm{H_3}^+$, HCO$^+$, H$^+$, and $\rm{N_2H^+}$; see reactions 17-34); (b) photodissociation by the interstellar radiation field (reactions 35-38); and (c) cosmic-ray-induced photoreactions (reactions 39-42). 
The rates of the destruction reactions with the abundant ions are estimated based on the approximated expression given in \citep{mill22}. The total dipole moment required for this calculation is available in the CDMS database \citep{mull16}. It is 1.67 Debye and 2.07 Debye for HC(S)SH and HC(O)SH, respectively. For the ion neutral destruction of CSSH and COSH, we consider the rate constants noted in Table \ref{table:reactions}.

\paragraph{Grain-surface Chemistry and Binding Energies}

For the ice phase formation of HC(S)SH, a radical-radical reaction between CS and HS followed by a hydrogenation reaction (grain surface reaction numbers 1 and  7) is proposed by \citet{lass19}. We consider a similar reaction between CO and HS followed by the hydrogenation reaction for forming ice phase HC(O)SH (grain surface reaction numbers  3 and 10) in our network. To understand the destruction of these species during the ice phase, we examine both thermal and non-thermal desorption processes, including reactive desorption (a fiducial factor of  0.03 is considered, following \citet{garr07}), photodesorption (by UV and external radiation field, as proposed in \citet{oberg09}), and cosmic-ray-induced desorption (as described in \citet{hase93}), as well as dissociation processes such as photodissociation and cosmic-ray-induced dissociation.

Appropriate binding energies (BEs) are essential for modeling these surface processes. As experimental data are lacking for these specific species, we performed quantum chemical calculations using the Gaussian 09 software \citep{fris13}. Recognizing that grain mantles in dense clouds are often water-rich, we calculated the BEs on a water-ice substrate, a standard and effective approach \citep{wake17,das18,sil24}.  The BEs of the studied species were computed using second-order M{\o}ller--Plesset perturbation theory (MP2) in conjunction with the aug-cc-pVDZ basis set. This level of theory provides a reliable description of electron correlation and is well suited for capturing weak intermolecular interactions such as van der Waals forces and hydrogen bonding. All geometries of the isolated species and the corresponding complexes were fully optimized without symmetry constraints, and harmonic vibrational frequency calculations were performed to confirm the nature of the stationary points as true minima. The binding energy was calculated using the supermolecular approach as $BE = E_{\mathrm{complex}} - (E_{\mathrm{surface}} + E_{\mathrm{adsorbate}})$, where $E_{\mathrm{complex}}$, $E_{\mathrm{surface}}$, and $E_{\mathrm{adsorbate}}$ represent the total energies of the optimized complex, surface model, and isolated molecule, respectively. 
The Zero-point energy (ZPE) corrections were included from harmonic frequency calculations at the same level of theory. 
To simplify the computation, we used a single water monomer and applied a scaling factor of 1.4 to approximate an amorphous water surface, as suggested by \citet{das18,wake17}. 
We derived the following BEs for our model: 3798\,K for \ch{HC(S)SH}, 3999\,K for \ch{HC(O)SH}, 3614\,K for CSSH, and 2412\,K for COSH. These values are broadly consistent with, but refined from, the values of 3722\,K, 2972\,K, and 4250\,K used by \citet{lass19} for \ch{HC(S)SH}, \ch{HC(O)SH}, and CSSH, respectively.

To properly model non-thermal desorption pathways, we also calculated the enthalpy of formation. 
 
The reaction enthalpies were computed using the standard thermochemical relation,
\begin{align*}
    \Delta_rH^\circ (298 \, \mathrm{K}) &= \sum_{\mathrm{products}} \Delta_fH^\circ_{\mathrm{prod}} (298 \, \mathrm{K}) \\
    &\quad - \sum_{\mathrm{reactants}} \Delta_fH^\circ_{\mathrm{react}} (298 \, \mathrm{K}).
\end{align*}

The enthalpies of formation were derived from atomization energies obtained at the DFT level employing the B3LYP/6-311+G(d,p) basis set. Experimental atomic heats of formation were adopted from \citet{curt97}. 
The resulting standard enthalpies of formation [$\Delta_f H^\circ (298 \, \mathrm{K})$] for the species  
used in our model were 156.8 kJ/mol for \ch{HC(S)SH}, -93.05 kJ/mol for \ch{HC(O)SH}, 304.98 kJ/mol for CSSH, and 44.4 kJ/mol for COSH.

\subsection{Modeling Results} \label{subsec:modeling_results}
We present the results of our chemical modeling, starting with a benchmark against a dark cloud environment before discussing the full hot corino simulation. The initial elemental abundances used for all models are listed in Table~\ref{tab:init}.

\begin{table}[]
    \centering
        \caption{  Initial elemental abundance taken from \cite{Esplugues2013}.}
    \begin{tabular}{c|c}
    \hline
    \hline
        He& 0.09\\
        O& $2.4 \times 10^{-4}$\\
        C$^+$& $1.7 \times 10^{-4}$\\
        N& $6.2 \times 10^{-5}$\\
        S$^+$& $1.5 \times 10^{-6}$\\
        Na$^+$&$2.0 \times 10^{-9}$ \\
        Mg$^+$& $7.0 \times 10^{-9}$\\
        Si$^+$& $8.0 \times 10^{-9}$\\
        P$^+$& $2.0 \times 10^{-10}$\\
        Cl$^+$& $1.0 \times 10^{-9}$\\
        Fe$^+$& $3.0 \times 10^{-9}$\\
        F$^+$& $6.7 \times 10^{-9}$\\

         \hline
    \end{tabular}
    \label{tab:init}
\end{table}

\begin{figure*}
    \centering
\includegraphics[width=1.0\linewidth]{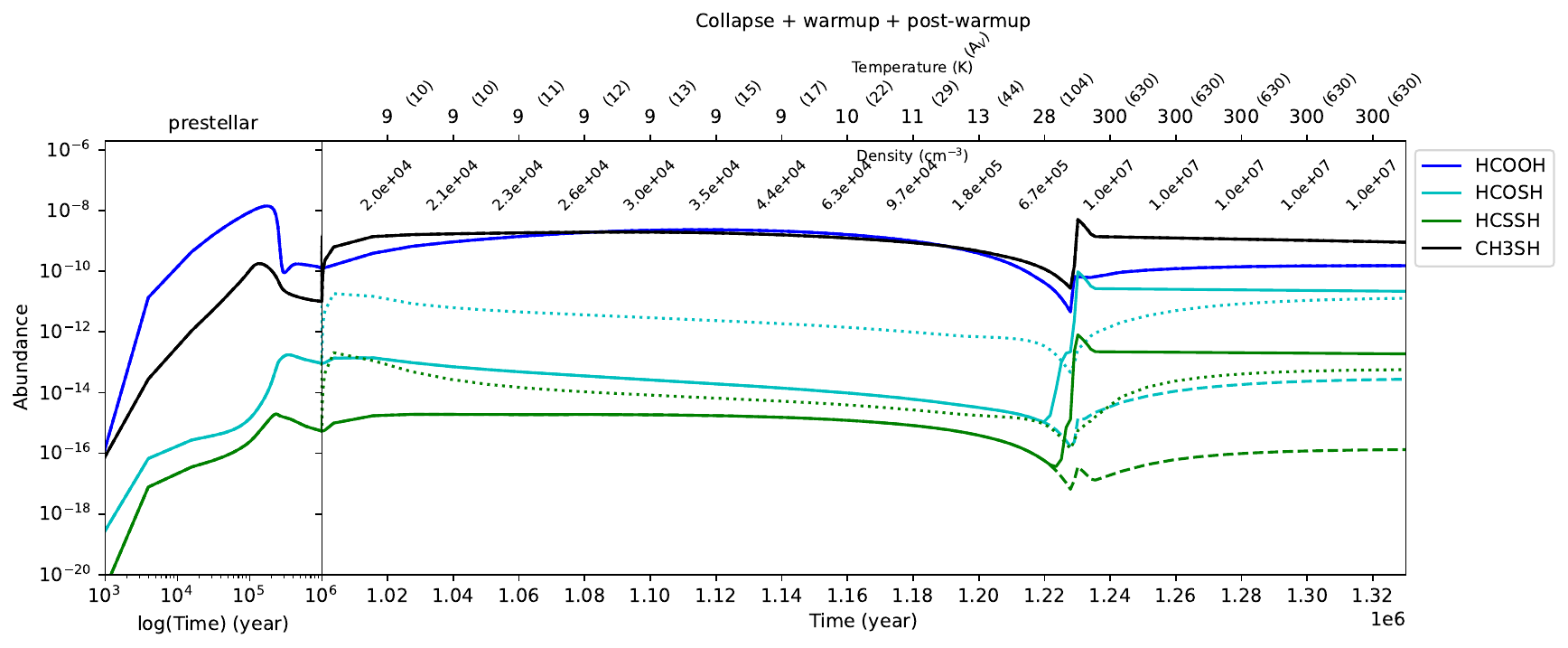}
    \caption {Time evolution of the fractional abundances of HCOOH, HC(O)SH, HC(S)SH, and CH$_3$SH. Solid lines represent the fiducial model, including all grain-surface and gas-phase reactions. Dashed lines show the abundances obtained when grain-surface reactions~5 and~6 are switched off, adopting realistic rate coefficients for gas-phase reactions~1 and~2. Dotted lines correspond to models in which grain-surface reactions~5 and~6 are excluded and the upper limits of the rate coefficients for gas-phase reactions~1 and~2 are adopted.}
    \label{fig:hcssh}
\end{figure*}

\paragraph{Benchmarking the model: }
As an initial step to validate our chemical network, we modeled a representative dark-cloud environment by considering three physically static evolutionary stages—diffuse, translucent, and dark cloud—following the approach adopted by \citet{lass19}.
 
Under these conditions, we find the following gas phase abundances: \(6.5 \times 10^{-11}\) for HCOOH, \(1.9 \times 10^{-11}\) for HC(O)SH, and \(1.04 \times 10^{-14}\) for HC(S)SH. These values are obtained by using the upper limit of the rate constants (\(\alpha = 10^{-10}\)) for gas phase reactions 1-2, as described in Table \ref{table:reactions}. \cite{lass19} obtained a 
 gas-phase fractional
abundance of $10^{-13}$.
The formation of these species in a dark cloud environment is primarily influenced by the gas-phase reactions 1, 2, 7, and 9 listed in Table \ref{table:reactions}.
 In contrast, when using more realistic reaction rates for the gas phase reactions 1 and 2, the resulting abundances of HC(O)SH and HC(S)SH are negligible. These values illustrate that HC(O)SH—and especially HC(S)SH—remain intrinsically scarce under dark‑cloud conditions.
 Due to the low temperature, the process of forming these species on the grain surface is found to be inefficient.

\paragraph{Hot Corino Model: 
}

 The solid lines of Figure~\ref{fig:hcssh} shows that below 40 K, the abundance of HC(S)SH and HC(O)SH are primarily contributed by reaction 7 and 9, respectively. At temperatures of $\sim$40–70~K. HC(O)SH is produced via the reaction between HCO and HS (grain-surface reaction~5), while HC(S)SH forms through the reaction between HCS and HS (grain-surface reaction~6). Using realistic gas-phase rate coefficients for reactions~1 and~2, the peak fractional abundances obtained with this are $2.3\times10^{-9}$ for HCOOH, $9.7\times10^{-11}$ for HC(O)SH, $8.05\times10^{-13}$ for HC(S)SH, and $5.1\times10^{-9}$ for CH$_3$SH. These values correspond to abundance ratios of $0.02:0.0002:1$ for HC(O)SH, HC(S)SH, and CH$_3$SH, respectively. 
We did not perform any transition state calculations for the ice phase reactions 5 and 6; thus, the abundances obtained using these reactions should be considered an upper limit.
In the absence of grain-surface reactions~5 and~6, and when adopting realistic rate coefficients for gas-phase reactions~1 and~2, the peak abundances of HC(O)SH and HC(S)SH decrease dramatically to $1.4\times10^{-13}$ and $1.9\times10^{-15}$, respectively (see the dashed curves in Fig.~\ref{fig:hcssh}). Even when the upper limits of the rate coefficients for gas-phase reactions~1 and~2 are adopted, the resulting peak abundances increase only to $2.4\times10^{-11}$ for HC(O)SH and $2.6\times10^{-13}$ for HC(S)SH.

\paragraph{Summary of Theoretical Findings.}
Our theoretical models are unambiguous: under all plausible conditions, the peak gas-phase abundance of t-\ch{HC(S)SH} is predicted to be extremely low, ranging from $\sim 10^{-15}$ in our standard model (dashed curve in Fig. \ref{fig:hcssh}) to, at most, $\sim 10^{-12}$ (solid curve in Fig. \ref{fig:hcssh}) in the most optimistic scenario with enhanced surface chemistry. These theoretical results are fully consistent with our observational non-detection and our  derived upper limit of $\leq 1 \times 10^{-10}$. A detailed comparison with the conflicting results of \citet{Manna2024ESC.....8.2401M} is presented in the Discussion.

\section{Discussion}
Our comprehensive reanalysis of the ALMA data for NGC 1333 IRAS 4A2 demonstrates that the recent claim of a t-\ch{HC(S)SH} detection by \citet{Manna2024ESC.....8.2401M} is unsubstantiated. The evidence against this detection is twofold: the observational data itself does not support the claim, and the reported abundance is inconsistent with plausible chemical abundances and  detailed astrochemical models. The comprehensive analysis and modeling presented here not only refute the prior claim but also provide a rigorous framework to guide future searches for dithiol and other complex sulfur-bearing species.

\subsection{Observational Refutation of the t-HC(S)SH Claim}
The primary argument against the detection lies in the clear misidentification of spectral features due to line blending. Our analysis in Section 3 shows that all  five candidate transitions of t-\ch{HC(S)SH} are severely blended with, or are entirely attributable to, known and abundant hot corino species. Specifically:
\begin{itemize}
    \item The features at 350.499 GHz and 350.849 GHz are definitively identified as strong transitions of glycolaldehyde (\ch{CH2OHCHO}) and methyl formate (\ch{CH3OCHO}), respectively. In addition, the emission feature around 351.588 GHz arises mostly from a cold component of \ch{CH3CHO}. Our global fits for these molecules across the entire band leave no room for any significant contribution from t-\ch{HC(S)SH}.
    \item The remaining two features are partially blended. Since the five t-\ch{HC(S)SH} transitions are expected to have nearly identical line strengths, a single molecule cannot produce the observed mixture of weak and strong spectral features. This inconsistency is further highlighted by the dissimilar peak intensities reported by \citet{Manna2024ESC.....8.2401M} (ranging from $\sim$0.6 to 0.9\,K)  and is also closely reproduced in Fig.~\ref{fig:LTE_upperlimit}. Clearly, this contradicts the expectation of uniform line strength for these specific transitions.
\end{itemize}

 In addition to the inconsistent analysis, the abundance of the abundant species \ch{CH3SH} is reported by \citet{Manna2024ESC.....8.2401M} to be $4.8 \times 10^{-10}$; this is nearly an order of magnitude lower than their reported value of $2.5 \times 10^{-9}$ for t-\ch{HC(S)SH}. Again, this implies an unrealistically high abundance for the dithiol. In comparison, we observationally find the abundance of \ch{CH3SH} to be $7.9 \times 10^{-10}$, a tentative abundance of t-\ch{HC(O)SH} of $1.6 \times 10^{-10}$, and an upper limit for t-\ch{HC(S)SH} of $\sim 10^{-10}$. Despite previous reports of a compact emission region for complex organic molecules \citep{Sahu2019ApJ...872..196S}, the authors' avoidance of high-resolution observations ($\sim 0.2\arcsec$) in favor of coarser resolution ($\sim 0.55\arcsec$) is not justified, as emission from chemical species may be diluted. However, we have utilized both sets of data, and they clearly imply the same result: a confirmed signature of line blending. Even without robust chemical modeling analysis, the observational evidence definitively indicates a misidentification arising from line blending and improper analysis.


\subsection{Chemical Context and Theoretical Plausibility}
Beyond the observational evidence, the claimed abundance of $\sim 2.5 \times 10^{-9}$ for t-\ch{HC(S)SH} is chemically implausible, as demonstrated by our detailed modeling.

First, this high abundance is inconsistent with the well-documented depletion of sulfur in dense interstellar environments. As established by \citet{lass19}, over 99\% of sulfur is expected to be frozen out onto dust grains during molecular cloud evolution, incorporated into simple ices like \ch{H2S_{ice}}, \ch{OCS_{ice}}, and \ch{HCS_{ice}}. Consequently, the gas phase is dominated by simple S-bearing species (e.g., HS, H$_2$S, CS, SO), whose abundances are typically on the order of $10^{-9}$ or less. It logically follows that more complex molecules like t-\ch{HC(S)SH} should be significantly less abundant than these simple precursors.

 Second, our detailed astrochemical models confirm this chemical intuition. Even under the most favorable conditions with optimistic reaction rates, our models struggle to produce a gas-phase t-\ch{HC(S)SH} abundance greater than  $\sim 10^{-12}$. While we acknowledge that model predictions carry uncertainties, general chemical trends provide a robust constraint. For instance, methanol (\ch{CH3OH})—the oxygenated counterpart of methanethiol (\ch{CH3SH})—is typically 1--2 orders of magnitude more abundant than formic acid (\ch{HCOOH}), the oxygenated analog of \ch{HC(S)SH}. Our model, based on a well-established network and benchmarked sulfur chemistry, implies relative abundances of \ch{CH3SH}:\ch{HC(O)SH}:\ch{HC(S)SH} $\approx$ 1 : $2\times 10^{-2}$ : $2\times 10^{-4}$. Even if we conservatively assume that \ch{HC(S)SH} is only one order of magnitude less abundant than \ch{CH3SH} (rather than the predicted three to four orders), the resulting abundance would still be significantly lower than our observational upper limit of $\sim 10^{-10}$. Combining these observational and modeling results, the value of $2.5\times 10^{-9}$ reported in the previous study is not only inconsistent with our findings but appears chemically impossible.

Finally, the theoretical justification in the previous report \citep{Manna2024ESC.....8.2401M} is further undermined by a critical methodological flaw: they compare a predicted \textit{ice-phase} abundance from their model with a \textit{gas-phase} observational constraint derived from rotational spectroscopy. This is a fundamentally invalid comparison that renders their modeling conclusions inapplicable.
of t-\ch{HC(S)SH} is unsubstantiated.

\subsection{Implications for Future Searches of Complex Sulfur Species}
Beyond refuting a specific claim, our work provides a robust framework that will be beneficial for future efforts to detect dithiol and other complex sulfur-bearing molecules. The challenges highlighted by this study offer clear guidance for the astrochemistry community.

First, observationally, the successful identification of new complex species in line-rich hot corinos demands exceptional rigor in blend analysis.  Our analysis \citep[also see, ][]{Synder2005ApJ...619..914S, Xue2019ApJ...882..118X} demonstrates that relying on a small number of transitions, especially when they are spectrally close or possess similar excitation energies, is fraught with risk. Future searches should prioritize using broad-bandwidth observations that cover transitions with a wide range of upper-state energies, as this is the only reliable way to constrain rotational temperatures and disentangle emission from contaminants. Furthermore, the search for potential blending species must be comprehensive and not arbitrarily constrained by energy or line strength cutoffs (e.g., $E_u < 500$\,K or a minimum $A_{ij}$ value). Our derived upper limit of $\sim 10^{-10}$ confirms that the detection of t-\ch{HC(S)SH} will remain challenging, not only due to sensitivity limitations but primarily due to severe spectral crowding from more abundant and common hot corino species.

Second, theoretically, our work provides a benchmarked chemical model for sulfur chemistry in hot corinos. We have presented new, quantum-chemically derived binding energies and a detailed reaction network that has been validated against known species like \ch{HC(O)SH}. Future modeling efforts can build upon this foundation. Our results underscore the necessity of benchmarking any chemical model against the observed abundances of multiple related species before its predictions for a new molecule can be considered reliable. The fundamental error of comparing modeled ice-phase abundances to gas-phase observations, as seen in previous work, must be avoided.

Finally, our study as a whole serves as a methodological template: a comprehensive inventory of all potential contaminants must be established before a new species can be claimed. By combining careful observational analysis, validated chemical modeling, and a clear understanding of the chemical network, the community can move forward with greater confidence in the search for the missing pieces of the interstellar sulfur puzzle.

\section{Conclusion}
We have performed a rigorous re-evaluation of the claimed detection of t-\ch{HC(S)SH} toward the hot corino NGC 1333 IRAS 4A2. Our analysis, combining observational spectroscopy and theoretical modeling, leads to the following unambiguous conclusions:
\begin{itemize}
    \item The previously reported detection of t-\ch{HC(S)SH} is refuted. The candidate spectral features are demonstrably caused by line blending with known, abundant hot corino molecules, primarily glycolaldehyde, methyl formate, and  acetaldehyde.
  \item We find no credible evidence for the presence of t-\ch{HC(S)SH} in the ALMA data. Considering the uncertainties in the emission features, the possible upper limit corresponds to a fractional abundance of $\sim 10^{-10}$, which is about one order of magnitude lower than the previously reported value.
    \item Our detailed astrochemical models predict a peak gas-phase abundance for t-\ch{HC(S)SH} of  $\sim 10^{-12}$, which is consistent with our observational non-detection and several orders of magnitude lower than the previously claimed value.
  \item The tentative abundance of the related species \ch{HC(O)SH} ($\sim 1-3 \times 10^{-10}$) and the abundant sulfur species \ch{CH3SH} ($\sim 10^{-9}$) provide strong chemical context, reinforcing the expectation that t-\ch{HC(S)SH} should be a very low-abundance species whose spectral signatures are likely unidentifiable in the observed spectra.
\end{itemize}
This work underscores the critical importance of exhaustive line blending analysis in spectrally dense regions. The reliable identification of new interstellar molecules, especially complex sulfur species, requires a multi-faceted approach that combines robust observational evidence with sound chemical and theoretical modeling.


\begin{acknowledgments}

D.S. acknowledges the support from Ramanujan Fellowship (SERB, RJF/2021/000116) and PRL. A.D. acknowledges MPE, Germany. V.M.R. acknowledges support from the grant PID2022-136814NB-I00 by the Spanish Ministry of Science, Innovation and Universities/State Agency of Research MICIU/AEI/10.13039/501100011033 and by ERDF, UE;  the grant RYC2020-029387-I funded by MICIU/AEI/10.13039/501100011033 and by ``ESF, Investing in your future'', and from the Consejo Superior de Investigaciones Cient{\'i}ficas (CSIC) and the Centro de Astrobiolog{\'i}a (CAB) through the project 20225AT015 (Proyectos intramurales especiales del CSIC), and the grant CNS2023-144464 funded by MICIU/AEI/10.13039/501100011033 and by ``European Union NextGenerationEU/PRTR''. This paper makes use of the following ALMA data: ADS/JAO.ALMA\#2015.1.00147.S. ALMA is a partnership of ESO (representing its member states), NSF (USA), and NINS (Japan), together with NRC (Canada) and MoST and ASIAA (Taiwan) and KASI (Republic of Korea), in cooperation with the Republic of Chile. The Joint ALMA Observatory is operated by ESO, AUI/NRAO, and NAOJ.
\end{acknowledgments}
\clearpage



\onecolumngrid 

\appendix
\restartappendixnumbering
\section{Comprehensive Search for Potential Contaminants}
\label{app:searched_molecules}

To ensure a robust analysis and account for all possible sources of spectral contamination, we performed an exhaustive search for numerous molecular species beyond those included in our final combined model. This search utilized the CDMS and JPL spectroscopic databases (see Tab.~\ref{tab:line_search1} to Tab.~\ref{tab:line_search5} ).
 The quantum numbers in  aforementioned tables are retrieved directly from the Splatalogue database \citep{Remijan2007AAS...21113211R} which aggregates entries from the JPL \citep{Pickett1998JQSRT..60..883P} CDMS \citep{Muller2005JMoSt.742..215M} catalogs. Both catalogs encode transitions using the QNFMT scheme, where transitions of asymmetric rotor molecules (QNFMT=3) are labeled as $J_{K_a, K_c}$ (upper $\rightarrow$ lower state): $J$ is the total rotational angular momentum quantum number, $K_a$ and $K_c$ are its projections in the prolate and oblate symmetric-top limits, respectively. For vibrationally excited species (QNFMT =14), the vibrational quantum number $v$ (or $v_t$ for torsional modes) is appended alongside the rotational quantum numbers. The quantities $F_1$ and $F$ denote intermediate and total coupled angular momenta arising from nuclear hyperfine interactions and appear only when hyperfine structure is resolved; these are absent for most complex organic molecules tabulated here. Any apparent differences in quantum number notation between entries reflect the different QNFMT codes assigned in the JPL and CDMS databases and do not represent inconsistencies in the tabulated values.

For the molecules mentioned here, we found either negligible contribution to the observed emission features, or could not fully confirm their presence due to severe blending. The list of searched-for but unconfirmed species includes:
    Acetic acid (\ch{CH3COOH}), Cyanomethanimine (\ch{HNCHCN}),  Dihydroxyacetone (\ch{CH2OHCOCH2OH}),  Disulfur monoxide (\ch{S2O}),  Ethanethiol (\ch{C2H5SH}),
     Ethylenimine (\ch{c-C2H4NH}),
     Ethyl isocyanide (\ch{CH3CH2NC}),
     Hydroxyacetonitrile (\ch{HOCH2CN}),
     n-propanol (\ch{n-C3H7OH}),
    Propargyl cyanide (\ch{HCCCH2CN}),
   1,3-propanediol (\ch{CH2OHCH2CH2OH}),
    Vinyl cyanide (\ch{CH2CHCN} and its isotopologue \ch{CH2CHC^{15}N}), Deuterated formic acid (\ch{DCOOH}), Isotopologues of abundant species, such as \ch{^{13}CH2OHCHO}, \ch{CH3O^{13}CHO}, \ch{CH3OC^{13}HO}, and \ch{^{13}CH3CH2CN}.

The detailed search process strengthens our conclusion that the emission features near the reported t-\ch{HC(S)SH} frequencies are mostly explained by the molecules included in our final inventory, and not by other unidentified contaminants.

\subsection{Search for a chemically relevant species \ch{HCOSH}}
As presented in section~\ref{sec:hcosh}, here we noted the details of relevant transitions in Tab.~\ref{tab:cHC(O)SH}, Tab.~\ref{tab:tHC(O)SH}. 

\begin{table*}[h!]
\centering
\caption{Key transitions of c-\ch{HC(O)SH} within the observed band.}
\label{tab:cHC(O)SH}
\begin{tabular}{lcccc}
\hline \hline
Molecule & Frequency (GHz) & Transition & $\log(A_{ij})$ & $E_\mathrm{up}$ (K) \\
\hline
c-\ch{HC(O)SH} & 350.3286712 & $13(2,11) - 12(1,12)$ & -3.6 & 62.3 \\
c-\ch{HC(O)SH} & 350.9226119 & $37(3,35) - 37(2,36)$ & -3.2 & 420.4 \\
c-\ch{HC(O)SH} & 351.2186572 & $31(1,31) - 30(1,30)$ & -3.1 & 273.6 \\
\hline
\end{tabular}
\end{table*}

\begin{table*}[h!]
\centering
\caption{Key transitions of t-\ch{HC(O)SH} within the observed band.}
\label{tab:tHC(O)SH}
\begin{tabular}{lcccc}
\hline \hline
Molecule & Frequency (GHz) & Transition & $\log(A_{ij})$ & $E_\mathrm{up}$ (K) \\
\hline
t-\ch{HC(O)SH} & 350.9384878 & $30(9,22) - 29(9,21)$ & -3.4 & 479.1 \\
t-\ch{HC(O)SH} & 350.9547795 & $6(3,4) - 5(2,3)$ & -4.2 & 36.0 \\
t-\ch{HC(O)SH} & 350.9574958 & $30(8,23) - 29(8,22)$ & -3.4 & 433.4 \\
t-\ch{HC(O)SH} & 351.0075470 & $31(0,31) - 30(0,30)$ & -3.3 & 272.6 \\
t-\ch{HC(O)SH} & 351.0107765 & $30(7,24) - 29(7,23)$ & -3.7 & 393.1 \\
t-\ch{HC(O)SH} & 351.1004545 & $6(3,3) - 5(2,4)$ & -4.2 & 36.0 \\
t-\ch{HC(O)SH} & 351.1205215 & $30(6,25) - 29(6,24)$ & -3.3 & 358.1 \\
t-\ch{HC(O)SH} & 351.2900213 & $30(3,28) - 29(3,27)$ & -3.3 & 285.6 \\
t-\ch{HC(O)SH} & 351.3302462 & $30(5,26) - 29(5,25)$ & -3.3 & 328.6 \\
\hline
\end{tabular}
\end{table*}


\section{Chemical Reaction Network}
\label{app:reactions}

Table~\ref{table:reactions} lists the key grain-surface and gas-phase reactions added to our chemical network to model the chemistry of \ch{HC(S)SH} and its analogs. Reaction rates are calculated following \citet{hase92} for surface reactions and are based on data from \citet{mill22} for gas-phase reactions where available.

\begin{longtable*}{lllll}

\caption{Key reactions for the formation and destruction of HC(S)SH and related species.}

\label{table:reactions}\\
\hline
\hline
& Reaction type & Reaction & Rate & References \\
\hline
\endfirsthead
\multicolumn{5}{c}{continued} \\
\hline
\hline
& Reaction type & Reaction & Rate & References \\
\hline
\endhead
\hline
\endfoot
\hline
\multicolumn{5}{c}{Grain surface reactions} \\
\hline
      1. &&  CS + HS $\rightarrow$ CSSH  &-&$E_a=2500 K$** \\
        2. &&  \hskip 1.3cm $\rightarrow$ HCS + S  &-&This work ($E_a=1000 K$) \\
        3. & & CO + HS $\rightarrow$ COSH  &-&This work ($E_a=2500 K$)\\
        4. &&  \hskip 1.3cm $\rightarrow$ OCS + H  &&This work ($E_a=1000 K$) \\
        5. & & HCO + HS $\rightarrow$ HC(O)SH$^a$  &-&This work\\
        6. & & HCS + HS $\rightarrow$ HC(S)SH$^a$  &-&This work\\

        7. &  & H + CSSH $\rightarrow$ HC(S)SH (10\%) &-&\citet{lass19}\\
        8. &  & \hskip 1.5cm $\rightarrow$ HS + HCS (70 \%) &-&This work\\
        9. &  &  \hskip 1.5cm $\rightarrow$ H$_2$S+ CS (20 \%) &-&This work\\
        10. & &  H + COSH $\rightarrow$ HC(O)SH (10\%)&-&This work\\
        11. &  & \hskip 1.5cm  $\rightarrow$ H$_2$ + OCS (70 \%) &-&This work\\
        12. &  & \hskip 1.5cm  $\rightarrow$ H$_2$O+ CS (20 \%) &-&This work\\
        13. & &  H + HC(O)SH $\rightarrow$ H$_2$+ COSH  &-&This work ($E_a=6150 K$)\\
        14. & &  H + HC(S)SH $\rightarrow$ H$_2$+ CSSH &-&This work($E_a=6150 K$)\\
        15. & &  CH$_3$ + HC(S)SH $\rightarrow$ CH$_4$+ CSSH &-&This work\\
        16. & &  CH$_3$ + HC(O)SH $\rightarrow$ CH$_4$+ COSH &-&This work\\
        17. & &  CH$_3$O + HC(S)SH $\rightarrow$ CH$_3$OH+ CSSH &-&This work\\
        18. & &  CH$_3$O + HC(O)SH $\rightarrow$ CH$_3$OH+ COSH &-&This work\\
        19. & &  NH$_2$ + HC(S)SH $\rightarrow$ NH$_3$+ CSSH &-&This work\\
        20. & &  NH$_2$ + HC(O)SH $\rightarrow$ NH$_3$+ COSH &-&This work\\
        21. & &  OH + HC(S)SH $\rightarrow$ H$_2$O+ CSSH &-&This work\\
        22. & &  OH + HC(O)SH $\rightarrow$ H$_2$O+ COSH &-&This work\\
        \hline
\multicolumn{5}{c}{Gas Phase Reactions} \\
\hline
1. & NN & $\mathrm{HS + H_2CS \rightarrow HC(S)SH + H}$ & $\alpha = 2.0 \times 10^{-13^*}, \, \beta = \gamma = 0$ & \citet{lass19} \\
2. & NN &  $\mathrm{ OH + H_2CS \rightarrow HC(O)SH + H}$ & $\alpha = 2.0 \times 10^{-13^*}, \, \beta = \gamma = 0$ & This work \\

3. & NN & $\mathrm{HS + H_2CS \rightarrow HCS + H_2S}$ & $\alpha = 7.73 \times 10^{-12}, \beta = -1.03, \gamma = 0$ & \citet{lass19} \\
4. & NN & $\mathrm{ OH + H_2CS \rightarrow HCS + H_2O}$ & $\alpha = 7.73 \times 10^{-12}, \beta = -1.03, \gamma = 0$ & This work \\
5. & RA & $\mathrm{H_2S + HCS^+ \rightarrow HC(S)SH_2^+ + PHOTON}$ & $\alpha = 4.0 \times 10^{-13}, \beta = -1.3, \gamma = 0$ & This work \\
6. & RA & $\mathrm{H_2S + HCO^+ \rightarrow HC(O)SH_2^+ + PHOTON}$ & $\alpha = 4.0 \times 10^{-13}, \beta = -1.3, \gamma = 0$ & This work \\
7. & DR & $\mathrm{HC(S)SH_2^+ + e^- \rightarrow HC(S)SH + H}$ & $\alpha = 1.1 \times 10^{-7}, \beta = -0.78, \gamma = 0$ & \citet{lass19} \\
8. & DR & $\mathrm{HC(S)SH_2^+ + e^- \rightarrow HCS + HS + H}$ & $\alpha = 7.33 \times 10^{-7}, \beta = -0.78, \gamma = 0$ & This work \\
9. & DR & $\mathrm{HC(O)SH_2^+ + e^- \rightarrow HC(O)SH + H}$ & $\alpha = 1.1 \times 10^{-7}, \beta = -0.78, \gamma = 0$ & This work \\
10. & DR & $\mathrm{HC(O)SH_2^+ + e^- \rightarrow HCO + HS + H}$ & $\alpha = 7.33 \times 10^{-7}, \beta = -0.78, \gamma = 0$ & This work \\
11. & DR & $\mathrm{HC(S)SH^+ + e^- \rightarrow HCS + HS}$ & $\alpha = 1.5 \times 10^{-7}, \beta = -0.5, \gamma = 0$ & This work \\
12. & DR & $\mathrm{HC(S)SH^+ + e^- \rightarrow CS + HS + H}$ & $\alpha = 1.5 \times 10^{-7}, \beta = -0.5, \gamma = 0$ & This work \\
13. & DR & $\mathrm{HC(O)SH^+ + e^- \rightarrow HCO + HS}$ & $\alpha = 1.5 \times 10^{-7}, \beta = -0.5, \gamma = 0$ & This work \\
14. & DR & $\mathrm{HC(O)SH^+ + e^- \rightarrow CO + HS + H}$ & $\alpha = 1.5 \times 10^{-7}, \beta = -0.5, \gamma = 0$ & This work \\
15. & IN & $\mathrm{H_2S + HCS^+ \rightarrow CS + H_3S^+}$ & \citet{mill22} & This work \\
16. & IN & $\mathrm{H_2S + HCO^+ \rightarrow CO + H_3S^+}$ & \citet{mill22} & This work \\
17. & IN & $\mathrm{H_3^+ + HC(S)SH \rightarrow H_3S^+ + CS + H_2}$ & \citet{mill22} & This work \\
18. & IN & $\mathrm{H_3^+ + HC(S)SH \rightarrow HCS^+ + H_2 + H_2S}$ & \citet{mill22} & This work \\
19. & IN & $\mathrm{H_3^+ + HC(O)SH \rightarrow H_3O^+ + CS + H_2}$ & \citet{mill22} & This work \\
20. & IN & $\mathrm{H_3^+ + HC(O)SH \rightarrow HCO^+ + H_2 + H_2S}$ & \citet{mill22} & This work \\
21. & IN & $\mathrm{H_3^+ + CSSH \rightarrow H_2S^+ + CS + H_2}$ & $\alpha = 1.8 \times 10^{-9}, \beta = -0.5, \gamma = 0$ & This work \\
22. & IN & $\mathrm{H_3^+ + CSSH \rightarrow HCS^+ + H_2 + HS}$ & $\alpha = 4.3 \times 10^{-9}, \beta = -0.5, \gamma = 0$ & This work \\
23. & IN & $\mathrm{H_3^+ + COSH \rightarrow H_3O^+ + CS + H}$ & $\alpha = 1.8 \times 10^{-9}, \beta = -0.5, \gamma = 0$ & This work \\
24. & IN & $\mathrm{H_3^+ + COSH \rightarrow HCO^+ + HS + H_2}$ & $\alpha = 4.3 \times 10^{-9}, \beta = -0.5, \gamma = 0$ & This work \\
25. & IN & $\mathrm{HCO^+ + HC(S)SH \rightarrow HCSH_2^+ + CO}$ & \citet{mill22} & This work \\
26. & IN & $\mathrm{HCO^+ + HC(O)SH \rightarrow HC(O)SH_2^+ + CO}$ & \citet{mill22} & This work \\
27. & IN & $\mathrm{HCO^+ + CSSH \rightarrow HCSH^+ + CO}$ & $\alpha = 1.8 \times 10^{-9}, \beta = -0.5, \gamma = 0$ & This work \\
28. & IN & $\mathrm{HCO^+ + COSH \rightarrow HC(O)SH^+ + CO}$ & $\alpha = 1.8 \times 10^{-9}, \beta = -0.5, \gamma = 0$ & This work \\
29. & CE & $\mathrm{H^+ + HC(S)SH \rightarrow HCSH^+ + H}$ & \citet{mill22} & This work \\
30. & CE & $\mathrm{H^+ + HC(O)SH \rightarrow HC(O)SH^+ + H}$ & \citet{mill22} & This work \\
31. & IN & $\mathrm{N_2H^+ + HC(S)SH \rightarrow HCSH_2^+ + N_2}$ & \citet{mill22} & This work \\
32. & IN & $\mathrm{N_2H^+ + HC(O)SH \rightarrow HC(O)SH_2^+ + N_2}$ & \citet{mill22} & This work \\
33. & IN & $\mathrm{N_2H^+ + CSSH \rightarrow HCSH^+ + N_2}$ & $\alpha = 1.32 \times 10^{-9}, \beta = -0.5, \gamma = 0$ & This work \\
34. & IN & $\mathrm{N_2H^+ + COSH \rightarrow HC(O)SH^+ + N_2}$ & $\alpha = 1.32 \times 10^{-9}, \beta = -0.5, \gamma = 0$ & This work \\
35. & PH & $\mathrm{HC(S)SH + PHOTON \rightarrow H_2S + CS}$ & $\alpha = 4.1 \times 10^{-10}, \beta = 0, \gamma = 1.8$ & This work \\
36. & PH & $\mathrm{HC(O)SH + PHOTON \rightarrow H_2S + CO}$ & $\alpha = 4.1 \times 10^{-10}, \beta = 0, \gamma = 1.8$ & This work \\
37. & PH & $\mathrm{CSSH + PHOTON \rightarrow HS + CS}$ & $\alpha = 4.1 \times 10^{-10}, \beta = 0, \gamma = 1.8$ & This work \\
38. & PH & $\mathrm{COSH + PHOTON \rightarrow HS + CO}$ & $\alpha = 4.1 \times 10^{-10}, \beta = 0, \gamma = 1.8$ & This work \\
39. & CR & $\mathrm{HC(S)SH + CRPHOT \rightarrow H_2S + CS}$ & $\alpha = 1.3 \times 10^{-17}, \beta = 0, \gamma = 124.5$ & This work \\
40. & CR & $\mathrm{HC(O)SH + CRPHOT \rightarrow H_2S + CO}$ & $\alpha = 1.3 \times 10^{-17}, \beta = 0, \gamma = 124.5$ & This work \\
41. & CR & $\mathrm{CSSH + CRPHOT \rightarrow HS + CS}$ & $\alpha = 1.3 \times 10^{-17}, \beta = 0, \gamma = 124.5$ & This work \\
42. & CR & $\mathrm{COSH + CRPHOT \rightarrow HS + CO}$ & $\alpha = 1.3 \times 10^{-17}, \beta = 0, \gamma = 124.5$ & This work \\
\hline

\multicolumn{5}{p{0.9\textwidth}}{\textit{Note.}---Reaction rate coefficients are given by the standard Arrhenius-Kooij formula $k = \alpha (T/300)^{\beta} \exp(-\gamma/T)$. For surface reactions, the rate is calculated following \citet{hase92}. Gas-phase reaction types are: NN (Neutral-Neutral), RA (Radiative Association), DR (Dissociative Recombination), IN (Ion-Neutral), CE (Charge Exchange), PH (Photodissociation), CR (Cosmic-Ray-induced Photodissociation). Binding Energies (BE) used in our model are: HC(S)SH = 3798\,K, CSSH = 3614\,K, HC(O)SH = 3999\,K, and COSH = 2412\,K. For comparison, \citet{lass19} used BEs of 3722\,K, 4250\,K, and 2972\,K for HC(S)SH, CSSH, and HC(O)SH, respectively.}\\
\multicolumn{5}{p{0.9\textwidth}}{\textit{$^a$ Not considered in default surface network}}\\
\multicolumn{5}{p{0.9\textwidth}}{\textit{$^*$An upper-limit of rate constant ($\alpha = 10^{-10}$) used to explore maximum production efficiency. $^{**}$\citet{lass19}}}

\end{longtable*}
\onecolumngrid 

\section{Derived Molecular Column Densities and Abundances}
\label{app:column_densities}

Table~\ref{tab:mcmc_set1} summarizes the best-fit parameters for the molecular species identified in this work. The column densities ($N_\mathrm{mol}$), excitation temperatures ($T_\mathrm{ex}$), line widths (FWHM), and source velocities ($V_\mathrm{LSR}$) are derived from the MCMC fitting procedure implemented in the CASSIS software package. 

 In the previous report, \citet{Manna2024ESC.....8.2401M} estimated the emission region size of t-\ch{HC(S)SH} to be $\sim 0\farcs66$ and mentioned that the emission is not resolved. Given this compact size, it is unclear why the authors opted to use coarser resolution data ($\sim 0\farcs55$) for their analysis, particularly when higher-resolution data ($\sim 0\farcs2$) with a maximum recoverable scale of $2\farcs44$ were available under the same observation ID. Consequently, for this reanalysis, we have utilized higher-resolution data to maximize the emission features for better identification, while also examining the coarser-resolution data to rigorously pinpoint line blending effects (see Fig.~\ref{fig:spectra2}).

The results of MCMC fitting for the higher-resolution data are presented in Table~\ref{tab:mcmc_set1}. It should be noted that due to severe line blending, the parameters for several molecules could not be fully constrained via MCMC fitting. In such cases, we adopted MCMC fitting as a basic gauze and considered typical hot corino abundances to estimate plausible values and generate representative synthetic spectra. To ensure reproducibility, the exact parameters used to generate these synthetic LTE spectra are listed in Table~\ref{tab:LTE_set1}. For completeness and comparison, the corresponding parameters derived from the coarser-resolution data are presented in Tables~\ref{tab:MCMC_set2} and \ref{tab:LTE_set2}.
\clearpage

\begin{deluxetable}{lccccc}[h!]
\vspace{-1cm}
\label{tab:mcmc_set1}
\tablecaption{MCMC-derived Parameters for Detected Molecular Species (resolution$\sim 0\farcs2$) }
\tablehead{
\colhead{Molecule} & \colhead{$N_\mathrm{mol}$} & \colhead{$T_\mathrm{ex}$} & \colhead{FWHM} & \colhead{$V_\mathrm{LSR}$} & \colhead{Abundance} \\
\colhead{} & \colhead{($10^{15}$\,\pcms)} & \colhead{(K)} & \colhead{(\kms)} & \colhead{(\kms)} & \colhead{($10^{-10}$)}
}
\startdata
\ch{c-HCOSH} & $12.44 \pm 0.89$ & $107.4 \pm 4.0$ & $2.98 \pm 0.14$ & $6.67 \pm 0.09$ & $4.98 \pm 0.36$ \\
\ch{t-HC(O)SH} & $4.02 \pm 0.31$ & $226.3 \pm 14.1$ & $2.14 \pm 0.04$ & $6.52 \pm 0.02$ & $1.61 \pm 0.12$ \\
\ch{CH3SH} & $19.77 \pm 0.99$ & $296.6 \pm 2.4$ & $2.41 \pm 0.08$ & $6.88 \pm 0.03$ & $7.91 \pm 0.40$ \\
\ch{HNCO} & $4.11 \pm 0.15$ & $164.6 \pm 5.2$ & $2.99 \pm 0.01$ & $7.00 \pm 0.03$ & $1.64 \pm 0.06$ \\
\ch{HN^{13}CO} & $2.11 \pm 0.13$ & $92.5 \pm 9.2$ & $2.99 \pm 0.01$ & $7.34 \pm 0.06$ & $0.84 \pm 0.05$ \\
\ch{a-CH_3CH_2OD} & $8.09 \pm 0.77$ & $106.9 \pm 11.0$ & $2.96 \pm 0.03$ & $6.52 \pm 0.01$ & $3.24 \pm 0.31$ \\
\ch{g'Ga-(CH_2OH)_2} & $41.58 \pm 6.96$ & $89.0 \pm 6.9$ & $2.97 \pm 0.04$ & $6.93 \pm 0.07$ & $16.63 \pm 2.78$ \\
\ch{CH_3CN} & $11.14 \pm 0.55$ & $211.1 \pm 4.5$ & $2.96 \pm 0.03$ & $6.91 \pm 0.02$ & $4.46 \pm 0.22$ \\
\ch{CH_3CHO} & $21.71 \pm 1.37$ & $228.1 \pm 15.1$ & $2.99 \pm 0.01$ & $6.65 \pm 0.04$ & $8.68 \pm 0.55$ \\
\ch{CH_3OCHO} & $86.60 \pm 3.04$ & $203.5 \pm 12.3$ & $2.88 \pm 0.09$ & $7.12 \pm 0.04$ & $34.64 \pm 1.22$ \\
\ch{C_2H_5OH} & $65.57 \pm 6.09$ & $148.4 \pm 16.9$ & $2.88 \pm 0.08$ & $7.18 \pm 0.06$ & $26.23 \pm 2.44$ \\
\ch{CH_2OHCHO} & $7.84 \pm 0.83$ & $150.0 \pm 24.6$ & $2.62 \pm 0.13$ & $7.02 \pm 0.09$ & $3.14 \pm 0.33$ \\
\ch{SO_2} & $10.88 \pm 1.50$ & $121.1 \pm 12.1$ & $2.98 \pm 0.02$ & $7.17 \pm 0.04$ & $4.35 \pm 0.60$ \\
\ch{CH_3COOH} & $19.08 \pm 1.50$ & $189.0 \pm 16.0$ & $1.80 \pm 0.04$ & $6.77 \pm 0.03$ & $7.63 \pm 0.60$ \\
\ch{^{13}CH_3OH} & $150.0 \pm 17.0$ & $274.9 \pm 21.0$ & $2.44 \pm 0.13$ & $6.79 \pm 0.04$ & $60.00 \pm 6.80$ \\
\enddata
\tablenotetext{}{Abundances are calculated with respect to H$_2$, assuming a total column density of $N_{\mathrm{H_2}} = 2.5 \times 10^{25}$\,\pcms. We used telescope ALMA with array size 1000m in the CASSIS MCMC fitting. }
\end{deluxetable}
-

\begin{deluxetable}{lccccc}[ht!]
\label{tab:LTE_set1}
\vspace{-1cm}
\tablecaption{Parameters for synthetic LTE spectra of individual species (resolution$\sim 0\farcs2$)}
\tablehead{
\colhead{Molecule} & \colhead{$N_\mathrm{mol}$} & \colhead{$T_\mathrm{ex}$} & \colhead{FWHM} & \colhead{$V_\mathrm{LSR}$} & \colhead{Size} \\
\colhead{} & \colhead{($10^{15}$\,\pcms)} & \colhead{(K)} & \colhead{(\kms)} & \colhead{(\kms)} & \colhead{(arcsec)}
}
\startdata
\ch{c-HC(O)SH}    & 12.8  & 109.5      & 2.99 & 6.5 & 0.68 \\
\ch{t-HC(O)SH}    & 7.0  & 300      & 2.99 & 6.65 & 0.68 \\
\ch{HNCO} & 3.8 & 161.8 & 7.03 & 0.96 \\
\ch{HN^{13}CO} & 1.9 & 80.7 & 2.99  & 7.4 & 0.76 \\
\ch{a-CH_3CH_2OD} & 7.48 & 80 & 2.98 & 6.5 & 0.96 \\
\ch{g'Ga-(CH_2OH)_2} & 40 & 84 & 2.99 &  7.0 & 1.8 \\
\ch{CH_3CN} & 6 & 218 &299 & 6.94 & 0.8 \\
\ch{CH_3CHO} & 25.7 & 240 & 2.98 & 6.65 & 0.56 \\
\ch{CH_3CHO} 2c(cold comp.) & 0.8 & 30 & 2.62 & 7 & 0.8\\
\ch{CH_3OCHO} & 86.6 & 232 & 2.7 & 7.12  & 1.0 \\
\ch{C_2H_5OH} & 67 & 200 & 2.8 & 7.2 & 0.8 \\
\ch{CH_2OHCHO} & 12 & 192 & 2.85 & 7.16 & 1.19 \\
\ch{SO_2} & $9.5 $ & 121 & 2.98 & 7.17 & 0.96 \\
\ch{CH_3COOH} &18 & 188 & 1.8 & 6.78 & 0.6 \\
\ch{^{13}CH_3OH} & 166 & 298 & 2.65 & 6.8 & 0.64 \\
\enddata
\end{deluxetable}

\begin{deluxetable}{lccccc}[ht!]
\label{tab:MCMC_set2}
\tablecaption{MCMC-derived Parameters for the confirmed blender of spectral features for the $0\farcs55$ resolution spectra}
\tablehead{
\colhead{Molecule} & \colhead{$N_\mathrm{mol}$} & \colhead{$T_\mathrm{ex}$} & \colhead{FWHM} & \colhead{$V_\mathrm{LSR}$} & \colhead{Abundance} \\
\colhead{} & \colhead{($10^{15}$\,\pcms)} & \colhead{(K)} & \colhead{(\kms)} & \colhead{(\kms)} & \colhead{($10^{-10}$)}
}
\startdata
\ch{g'Ga-(CH_2OH)_2} & $0.81 \pm 0.49$ & $92.4 \pm 18.9$ & $1.80 \pm 0.12$ & $7.12 \pm 0.06$ & $1.65 \pm 1.00$ \\
\ch{CH_2OHCHO} & $0.07 \pm 0.07$ & $280.5 \pm 16.0$ & $1.60 \pm 0.06$ & $6.60 \pm 0.04$ & $0.14 \pm 0.14$ \\
\ch{CH_3CHO} & $0.83 \pm 0.20$ & $120.7 \pm 11.1$ & $2.21 \pm 0.32$ & $7.14 \pm 0.11$ & $1.69 \pm 0.41$ \\
\ch{CH_3OCHO} & $6.77 \pm 1.06$ & $165.4 \pm 7.7$ & $1.91 \pm 0.14$ & $6.79 \pm 0.04$ & $13.82 \pm 2.16$ \\
\enddata
\tablenotetext{}{Abundances are calculated with respect to H$_2$, assuming a total column density of $N_{\mathrm{H_2}} = 4.9 \times 10^{24}$\,\pcms. We used telescope ALMA with array size 400m in the CASSIS MCMC fitting.}
\end{deluxetable}

\begin{deluxetable}{lccccc}[ht!]
\label{tab:LTE_set2}
\tablecaption{Parameter for synthetic LTE spectra used for coarser ($0\farcs55$) resolution spectra}
\tablehead{
\colhead{Molecule} & \colhead{$N_\mathrm{mol}$} & \colhead{$T_\mathrm{ex}$} & \colhead{FWHM} & \colhead{$V_\mathrm{LSR}$} & \colhead{Size} \\
\colhead{} & \colhead{($10^{15}$\,\pcms)} & \colhead{(K)} & \colhead{(\kms)} & \colhead{(\kms)} & \colhead{(arcsec)}
}
\startdata
    \ch{g'Ga-(CH_2OH)_2} & 0.4 & 100 & 2.5 & 6.8 & 1.4 \\
\ch{CH_2OHCHO} & 0.35 & 100 & 1.92 & 6.84 & 1.4 \\
\ch{CH_3CHO} & 0.9 & 104.8 & 2.62 & 7.0 & 0.8 \\
\ch{CH_3CHO}, cold comp. & 0.8 & 30 & 2.62 & 7.0 & 0.8 \\
\ch{CH_3OCHO} & 4.2 & 155 & 2.23 & 6.9 & 0.82 \\
\enddata
\end{deluxetable}

\onecolumngrid 
\section{Details of line analysis approach } \label{supporting: analysis}

\subsection{Search for the Possible Presence of Potential Species around the \ch{t-HC(S)SH} Transition Frequency}

To verify whether the spectral signature of \ch{HC(S)SH} is actually present towards NGC 1333 IRAS4A2, we checked the line transitions from the database (CDMS/JPL/Splatalogue). Here we just note the transitions with $\pm 4$ km/s around the \ch{t-HC(S)SH} transition with $ \text{A}_{ij,\text{min}} = -5 $ and $ E_u < 500 \, \text{K} $, but note that we have not limited our search only for these set of lines. For example, the \ch{CH3CN} higher excitation lines have strong presence in the spectra, but not present in the example tables.  We have manually removed multiple transitions for the species \ch{ClONO2}, as it is an atmospheric species. Additionally, we have excluded several transitions of \ch{NH2CH2CH2OH} v$_{25}$
=1, as such high vibrational transitions are unlikely in this environment. These steps were taken to keep the table concise and clear; however, our broader search philosophy is designed to remain unbiased and is not limited to the subset of lines shown here.

In the table, \ch{t-HC(S)SH} is marked in bold for better visibility, and potential weeds or blended transitions can be qualitatively found by looking at  high $ \text{A}_{ij} $ values and common molecules that are previously reported in hot corions. The possible molecules that can be present within the searched range were further modeled using the CASSIS software Line Analysis module to confirm or discard their presence.

\begin{table*}
\caption{Transitions around t-HC(S)SH 54(1,54)-53(1,53)}
\begin{tabular}{|lcccc|}
\hline
Species Name & Quantum Number Transition & Frequency (GHz) & E$_u$ (K) & $\log_{10}(A_{ij} / \text{s}^{-1})$ \\
\hline
\ch{^{37}ClONO2} & 22(13,9)-21(12,10),J=41/2-39/2 & 350.3479 & 139.0 & -4.9 \\
\ch{NH2CO2CH3} v=0 & 45(10,36)-44(9,35)A & 350.3488 & 420.1 & -2.9 \\
\ch{CH2OH13CHO} & 30(9,22)-29(9,21) & 350.3493 & 306.6 & -4.8 \\
\ch{CH3^{13}CHO,} vt le; 1 & 16(8,9)-17(7,11)E,vt=1 & 350.3495 & 468.5 & -4.6 \\
\ch{Z-HNCHCN} & 36(13,23)-35(13,22) & 350.3503 & 706.7 & -3.4 \\
\ch{Z-HNCHCN} & 36(13,24)-35(13,23) & 350.3503 & 706.7 & -3.4 \\
\ch{g-C2H5^{34}SH} & 43(8,36)-43(7,36),v= 0-1 & 350.3503 & 523.6 & -4.4 \\
\ch{{\bf trans-HC(S)SH}} & 54(1,54)-53(1,53) & 350.3516 & 466.4 & -3.3 \\
\ch{cis-CH2OHCHO} v=1 & 29(7,22)-28(7,21) & 350.3536 & 555.4 & -4.8 \\
\ch{c-C2H4NH} & 8(4,5)-7(3,5) & 350.3546 & 65.5 & -3.6 \\
\ch{g-n-C3H7CN,} v30 = 1 & 58(10,49)-57(10,48) & 350.3549 & 696.4 & -2.6 \\
\ch{OS^{17}O} & 26(9,18)-27(8,19) & 350.3555 & 526.2 & -2.8 \\
\ch{OS^{17}O} & 26(9,17)-27(8,20) & 350.3556 & 526.2 & -2.8 \\
\ch{Ga-n-C3H7OH} & 37(6,32)-36(6,31) & 350.3558 & 338.0 & -4.2 \\
\hline
\end{tabular}
\label{tab:line_search1}
\end{table*}

\begin{table*}
\caption{Transitions around t-HC(S)SH  52(3,49)-51(3,48)}
\begin{tabular}{|lcccc|}
\hline
Species Name & Quantum Number Transition & Frequency (GHz) & E$_u$ (K) & $\log_{10}(A_{ij} / \text{s}^{-1})$ \\
\hline
\ch{Ga-n-C3H7OH} & 37(18,20)-36(18,19) & 350.3969 & 468.4 & -4.3 \\
\ch{Ga-n-C3H7OH} & 37(18,19)-36(18,18) & 350.3969 & 468.4 & -4.3 \\
\ch{NH2CH2CH2OH} v25=1 & 15(11,4)-14(10,4) & 350.3971 & 552.1 & -4.5 \\
\ch{CH2F2} v=0 & 15(5,10)-15(4,11) & 350.3973 & 161.6 & -3.3 \\
\ch{C3H6O2} & 52(30,23)-51(30,22)E & 350.3978 & 734.9 & -3.3 \\
\ch{TiO2} & 26(0,26)-25(1,25) & 350.3989 & 230.9 & -2.0 \\
\ch{HCCCH2OH} & 31(10,21)-30(11,19),vt= 1-0 & 350.3991 & 378.6 & -4.8 \\
\ch{O^{79}BrO} & 29(20,9)-30(19,12),J=57/2-59/2,F=29-30 & 350.4000 & 698.8 & -4.2 \\
\ch{^{37}ClONO2} & 76(3,73)-75(4,72),J=151/2-149/2 & 350.4003 & 672.8 & -4.8 \\
\ch{{\bf trans-HC(S)SH}} & 52(3,49)-51(3,48) & 350.4011 & 463.0 & -3.3 \\
\ch{ClOOCl} & 80(6,74)-79(7,73) & 350.4015 & 733.9 & -4.3 \\
\ch{g'Ga-(CH2OH)2} & 37(1,36)v= 0-36(2,35)v= 0 & 350.4017 & 327.7 & -3.7 \\
\ch{g'Ga-(CH2OH)2} & 37(2,36)v= 0-36(1,35)v= 0 & 350.4017 & 327.7 & -3.7 \\
\ch{a-n-C3H7CN,} v30 = 1 & 46(4,42)-45(3,43) & 350.4018 & 445.8 & -4.2 \\
\ch{CH2CHC^{15}N} & 38(11,27)-37(11,26) & 350.4020 & 587.7 & -2.5 \\
\ch{CH2CHC^{15}N} & 38(11,28)-37(11,27) & 350.4020 & 587.7 & -2.5 \\
\ch{CH3O^{13}CHO,} vt = 0, 1 & 29(22,8)-28(22,7),v = 4-4 & 350.4032 & 762.7 & -3.5 \\
\ch{S2O} v = 0 & 43(1,42)-43(0,43) & 350.4040 & 434.9 & -4.2 \\
\ch{c-C4H4O2} & 70(6,64)-69(6,63) & 350.4041 & 631.8 & -2.9 \\
\ch{c-C6H5COCH3} & 41(33,8)-40(32,9) & 350.4048 & 305.1 & -3.7 \\
\ch{c-C6H5COCH3} & 41(33,9)-40(32,8) & 350.4048 & 305.1 & -3.7 \\
\ch{HOCH2CN} & 38(11,27)-37(11,26),v = 1-1 & 350.4052 & 501.3 & -2.9 \\
\ch{HOCH2CN} & 38(11,28)-37(11,27),v = 1-1 & 350.4052 & 501.3 & -2.9 \\
\hline
\end{tabular}
\label{tab:trasition2}
\end{table*}


\begin{table*}
\caption{Transitions around t-HC(S)SH 54(0,54)-53(0,53)}
\label{tab:transition3}
\begin{tabular}{|lcccc|}
\hline 
Species Name & Quantum Number Transition & Frequency (GHz) & E$_u$ (K) & $\log_{10}(A_{ij} / \text{s}^{-1})$ \\
\hline
\ch{CH3COOH} v=0 & 23(10,13)-22(11,12),E & 350.4954 & 232.2 & -3.6 \\
\ch{c-C3H5CN} & 52(24,28)-51(24,27) & 350.4963 & 788.5 & -2.5 \\
\ch{S4} & 46(11,35)-45(10,36) & 350.4972 & 219.3 & -4.9 \\
\ch{cis-CH2OHCHO} v=0 & 30(11,19)-29(11,18) & 350.4982 & 332.6 & -4.8 \\
\ch{i-C3H7CN} & 50(25,25)-49(25,24) & 350.4983 & 559.7 & -2.5 \\
\ch{i-C3H7CN} & 50(25,26)-49(25,25) & 350.4983 & 559.7 & -2.5 \\
\ch{g'G'g-CH3CHOHCH2OH} & 53(15,39)-52(15,38) & 350.4995 & 504.1 & -4.4 \\
\ch{\bf{trans-HC(S)SH}} & 54(0,54)-53(0,53) & 350.4996 & 466.4 & -3.3 \\
\ch{NH2CO2CH3} v=1 & 50(6,45)-49(6,44)A & 350.5001 & 634.2 & -4.2 \\
\ch{cis-CH2OHCHO} v=0 & 33(2,31)-32(3,30) & 350.5002 & 301.7 & -3.0 \\
\ch{CH3COOH} v=0 & 23(11,13)-22(11,12),E & 350.5005 & 232.2 & -3.9 \\
\ch{^{46}TiO2} & 13(8,6)-13(7,7) & 350.5007 & 138.2 & -2.4 \\
\ch{HOCH2CN} & 38(8,31)-37(8,30),v = 1-1 & 350.5007 & 422.3 & -2.9 \\
\ch{HOCH2CN} & 38(8,30)-37(8,29),v = 1-1 & 350.5008 & 422.3 & -2.9 \\
\ch{C2H5CHCNCH3} & 65(30,35)-64(31,34) & 350.5012 & 626.0 & -3.0 \\
\ch{cis-CH2OHCHO} v=0 & 33(3,31)-32(3,30) & 350.5014 & 301.7 & -4.8 \\
\ch{NH2CO2CH3} v=1 & 48(8,41)-47(8,40)A & 350.5015 & 622.7 & -4.3 \\
\ch{C3H6O2} & 71(2,69)-71(1,70)E & 350.5018 & 748.2 & -3.9 \\
\ch{C3H6O2} & 71(3,69)-71(2,70)E & 350.5019 & 748.2 & -3.9 \\
\ch{cis-CH2OHCHO} v=0 & 33(2,31)-32(2,30) & 350.5021 & 301.7 & -4.8 \\
\ch{cis-CH2OHCHO} v=0 & 33(3,31)-32(2,30) & 350.5031 & 301.7 & -3.0 \\
\ch{CH3OCHO} v=1 & 17(13,5)-17(12,6)E & 350.5035 & 390.1 & -4.5 \\
\ch{c-C4H4O2} & 68(14,55)-67(14,54) & 350.5040 & 666.4 & -2.9 \\
\hline

\hline
\end{tabular}
\end{table*}


\begin{table*}
\caption{Transitions around t-HC(S)SH 52(2,50)-51(2,49)}
\begin{tabular}{|lcccc|}
\hline
Species Name & Quantum Number Transition & Frequency (GHz) & E$_u$(K) & $\log_{10}(A_{ij} / \text{s}^{-1})$ \\
\hline
\ch{CH3OCHO} v=1 & 31(3,29)-30(3,28)A & 350.8449 & 472.3 & -3.2 \\
\ch{OC(CN)2} & 46(43,3)-46(42,4) & 350.8454 & 632.6 & -4.9 \\
\ch{OC(CN)2} & 46(43,4)-46(42,5) & 350.8454 & 632.6 & -4.9 \\
\ch{gG'a-CH3CHOHCH2OH} & 28(17,11)-27(16,11) & 350.8461 & 197.7 & -4.5 \\
\ch{gG'a-CH3CHOHCH2OH} & 28(17,11)-27(16,12) & 350.8461 & 197.7 & -4.9 \\
\ch{gG'a-CH3CHOHCH2OH} & 28(17,12)-27(16,11) & 350.8461 & 197.7 & -4.9 \\
\ch{gG'a-CH3CHOHCH2OH} & 28(17,12)-27(16,12) & 350.8461 & 197.7 & -4.5 \\
\ch{c-^{13}CCCH} & 18(8,10)-18(8,11),J=37/2-35/2,  & 350.8469 & 518.3 & -3.8 \\
 &  F1=18-18,F=35/2-35/2 &  &  &  \\
\ch{NH2CO2CH3} v=0 & 45(11,35)-44(11,34)A & 350.8472 & 427.7 & -4.4 \\
\ch{H^{18}ONO2} & 18(15,3)-17(15,2) & 350.8478 & 194.3 & -3.6 \\
\ch{c-^{13}CCCH} & 18(8,10)-18(8,11),J=37/2-35/2, & 350.8491 & 518.3 & -4.2 \\
    & F1=19-18,F=37/2-37/2 &   &   &   \\
\ch{\bf{trans-HC(S)SH}} & 52(2,50)-51(2,49) & 350.8495 & 454.6 & -3.3 \\
\ch{H2NCH2COOH} - I v=0 & 22(15,7)-21(14,8) & 350.8495 & 157.7 & -4.2 \\
\ch{t-CH3CH2OH} & 33(1,32)-33(0,33) & 350.8498 & 463.4 & -3.3 \\
\ch{C2H5OH} & 33(1,32)-33(0,33),anti & 350.8498 & 463.4 & -4.1 \\
\ch{HCCCH2NH2} & 40(16,24)-39(16,23) & 350.8503 & 653.2 & -4.1 \\
\ch{HCCCH2NH2} & 40(16,25)-39(16,24) & 350.8503 & 653.2 & -4.1 \\
\ch{g'Ga-(CH2OH)2} & 34(18,16)v= 0-34(17,18)v= 1 & 350.8504 & 451.4 & -4.7 \\
\ch{CH3OCHO} v=1 & 31(2,29)-30(2,28)A & 350.8509 & 472.3 & -3.2 \\
\ch{c-C6H4} & 37(19,19)-36(18,18) & 350.8533 & 367.0 & -3.7 \\
\hline
\end{tabular}
\label{tab: transition4}
\end{table*}

\begin{table*} 
\caption{Transitions around t-HC(S)SH 53(1,52)-52(1,51) }
\begin{tabular}{|lcccc|}
\hline
Species Name & Quantum Number Transition & Frequency (GHz) & E$_u$(K) & $\log_{10}(A_{ij} / \text{s}^{-1})$ \\
\hline
\ch{^{13}CH2CHCN} & 38(6,33)-37(6,32) & 351.5837 & 405.7 & -2.5 \\
\ch{c-C3H5CN} & 32(6,26)-31(5,26) & 351.5840 & 192.6 & -4.1 \\
\ch{HOONO2} & 21(12,9)-20(11,9),v=1-0,vt(NO2)=1 & 351.5862 & 353.4 & -3.6 \\
\ch{HOONO2} & 21(12,10)-20(11,10),v=1-0,vt(NO2)=1 & 351.5862 & 353.4 & -3.6 \\
\ch{^{13}CH2CHCN} & 38(6,32)-37(6,31) & 351.5872 & 405.7 & -2.5 \\
\ch{c-HDC3O} & 36(3,33)-36(3,34) & 351.5878 & 443.0 & -4.1 \\
\ch{{\bf trans-HC(S)SH}} & 53(1,52)-52(1,51) & 351.5882 & 462.9 & -3.3 \\
\ch{C3H6O2} & 38(27,12)-38(26,12)E & 351.5887 & 474.6 & -3.3 \\
\ch{CH3CHO} v = 0, 1, 2 & 6(3,3)-5(2,4)A,vt=0 & 351.5889 & 39.8 & -3.9 \\
\ch{t-C4H9CN} & 64(-12)-63(12) & 351.5892 & 560.9 & -2.4 \\
\ch{t-C4H9CN} & 64(12)-63(-12) & 351.5892 & 560.9 & -2.4 \\
\ch{a-CH3^{13}CH2OH} & 39(3,36)-39(2,37) & 351.5893 & 671.6 & -3.7 \\
\ch{HCCCH2OD} & 40(16,24)-39(16,23),vt= 1-1 & 351.5903 & 664.9 & -3.7 \\
\ch{^{80}SeO2} & 26(19,7)-27(18,10) & 351.5907 & 622.2 & -4.3 \\
\ch{H^{18}ONO2} & 12(8,5)-11(5,6) & 351.5912 & 82.6 & -4.1 \\
\ch{CH3OCN} & 34(3,31)-33(3,30),E & 351.5915 & 307.3 & -2.4 \\
\hline
\end{tabular}
\label{tab:line_search5}
\end{table*}


\subsection{List of Confirmed Molecules and Blenders }

As mentioned in the previous section, based on that search, we found the confirmed presence of the molecules \ch{CH3OCHO},  \ch{CH2OHCHO} and \ch{CH3CHO}.  Additionally, \ch{gGa-(CH2OH)2} is partially blended with the second transition of t-HC(S)SH. For the first transition of t-HC(S)SH, we did not find any spectral features that could fully account for the emission profile. However, part of the emission appears to arise from \ch{CH3CN}, whose presence is confirmed through multiple non-blended transitions. Notably, \ch{CH3CN} transitions are not listed in the table, as they arise from the $v_8 = 1$ vibrational state and have upper-state energies ($E_u$) higher than those included in the table. Synthetic spectra of major molecules are shown in the figures.  

\begin{figure*}
    \centering
    \includegraphics[width=1.0\linewidth]{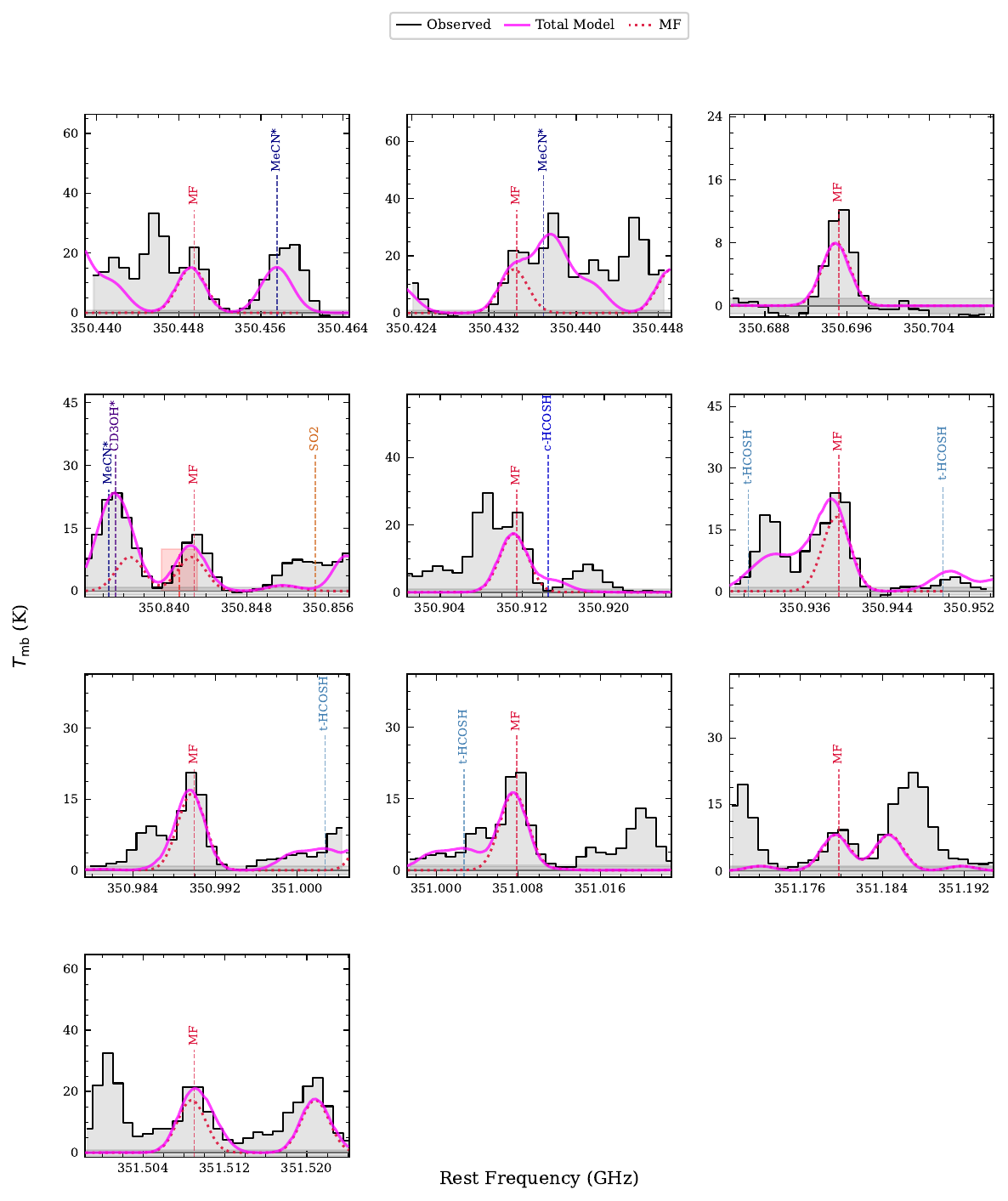}
    \caption{The observed high-resolution ALMA spectrum (black) is overplotted with the synthetic LTE model of \ch{CH_3OCHO} (dotted) in addition to total model (magenta) and . The shaded red region indicates the expected frequency range of the $t$-HC(S)SH transition assuming a $V_{LSR} = 6.9$ km s$^{-1}$ with a total width of 3 km s$^{-1}$, identical to Fig. 1. Only major transitions are displayed to illustrate the significant contribution of this species to the emission features near the $t$-HC(S)SH transition frequencies.}
    \label{fig:ch3ocho}
\end{figure*}

\begin{figure*}
    \centering
    \includegraphics[width=1.0\linewidth]{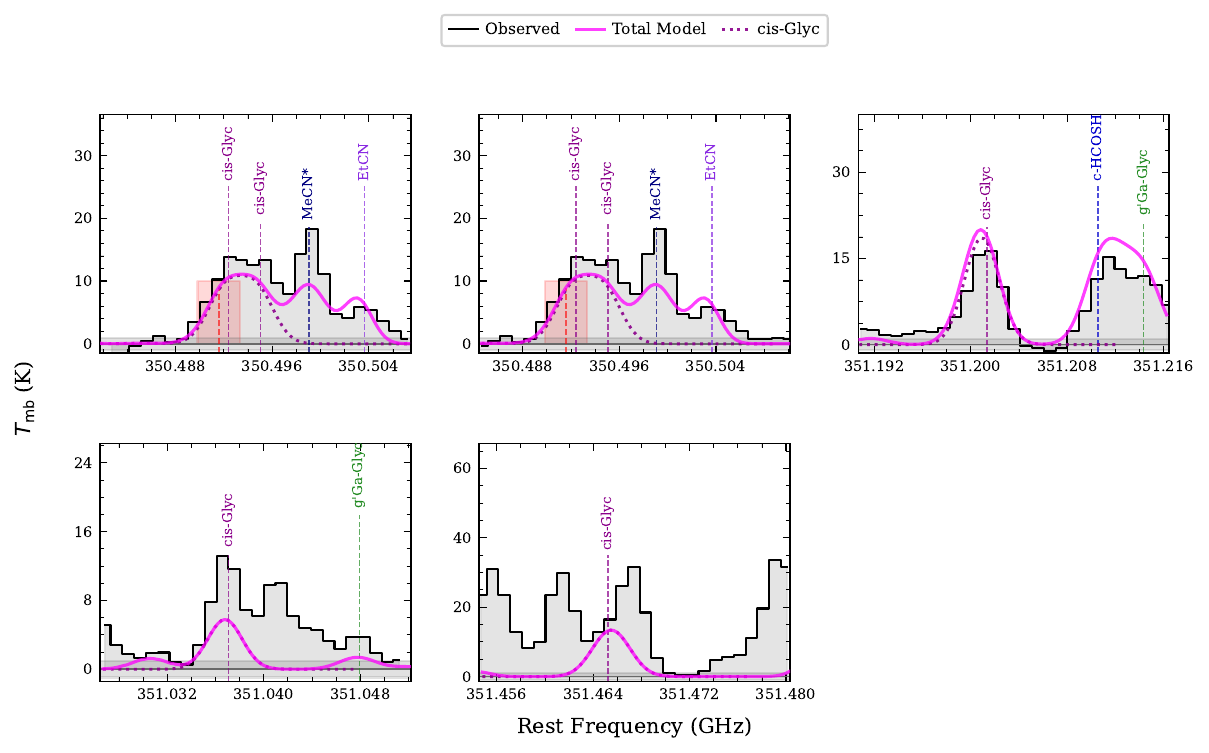}
    \caption{ Synthetic LTE spectra of Glycolaldehyde (\ch{CH_2OHCHO}). Same as Fig.~\ref{fig:ch3ocho} but for \ch{CH_2OHCHO}.}
    \label{fig: CH2OHCHO}
\end{figure*}

\begin{figure*}
    \centering
    \includegraphics[width=1.0\linewidth]{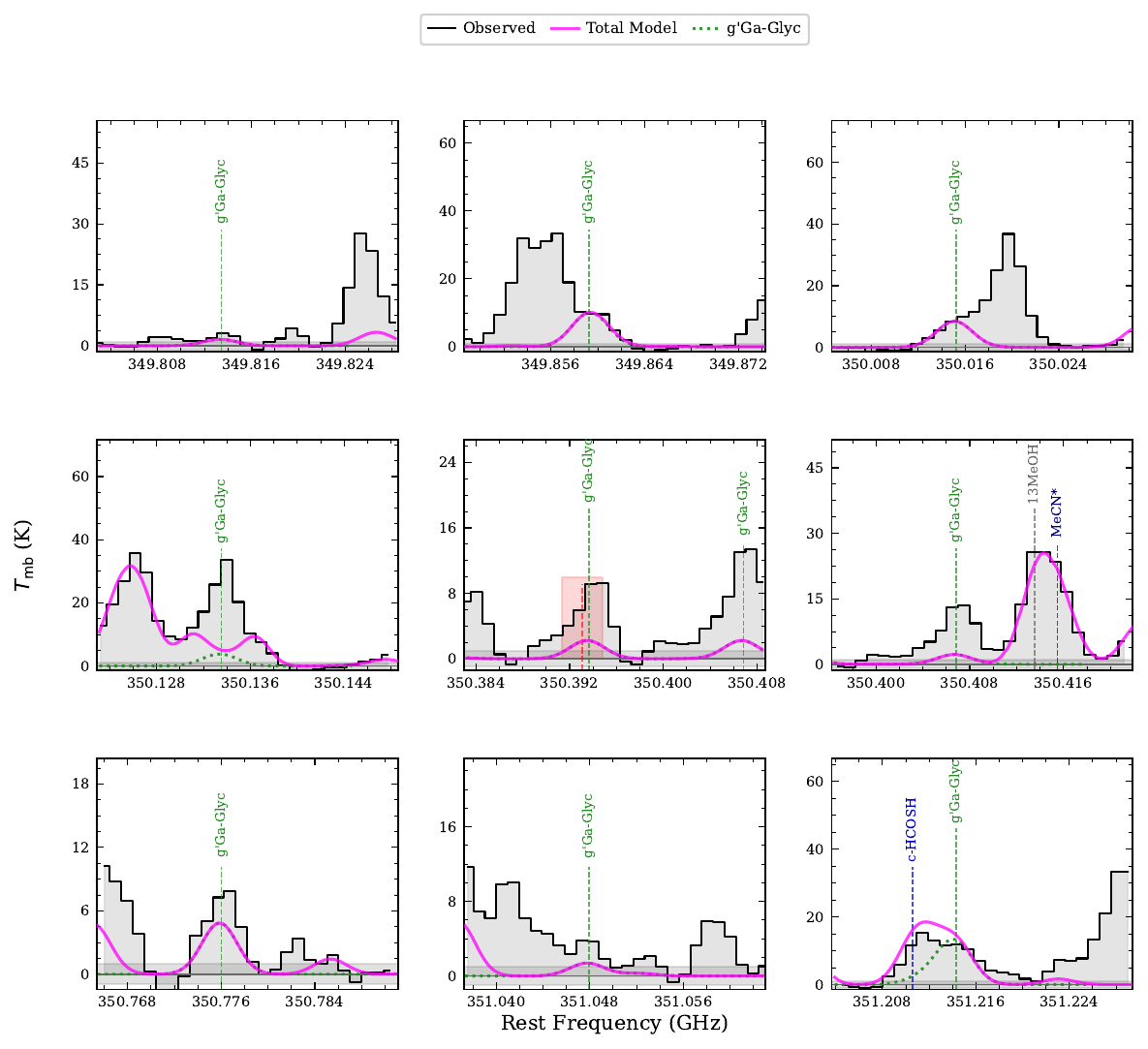}
    \caption{Synthetic LTE spectra of Ethylene Glycol ($g'Ga\text{-}(CH_2OH)_2$). Same as Fig.~\ref{fig:ch3ocho} but for $g'Ga\text{-}(CH_2OH)_2$.}
    \label{fig:ethylene glycol}
\end{figure*}

\begin{figure*}
    \centering
    \includegraphics[width=1.0\linewidth]{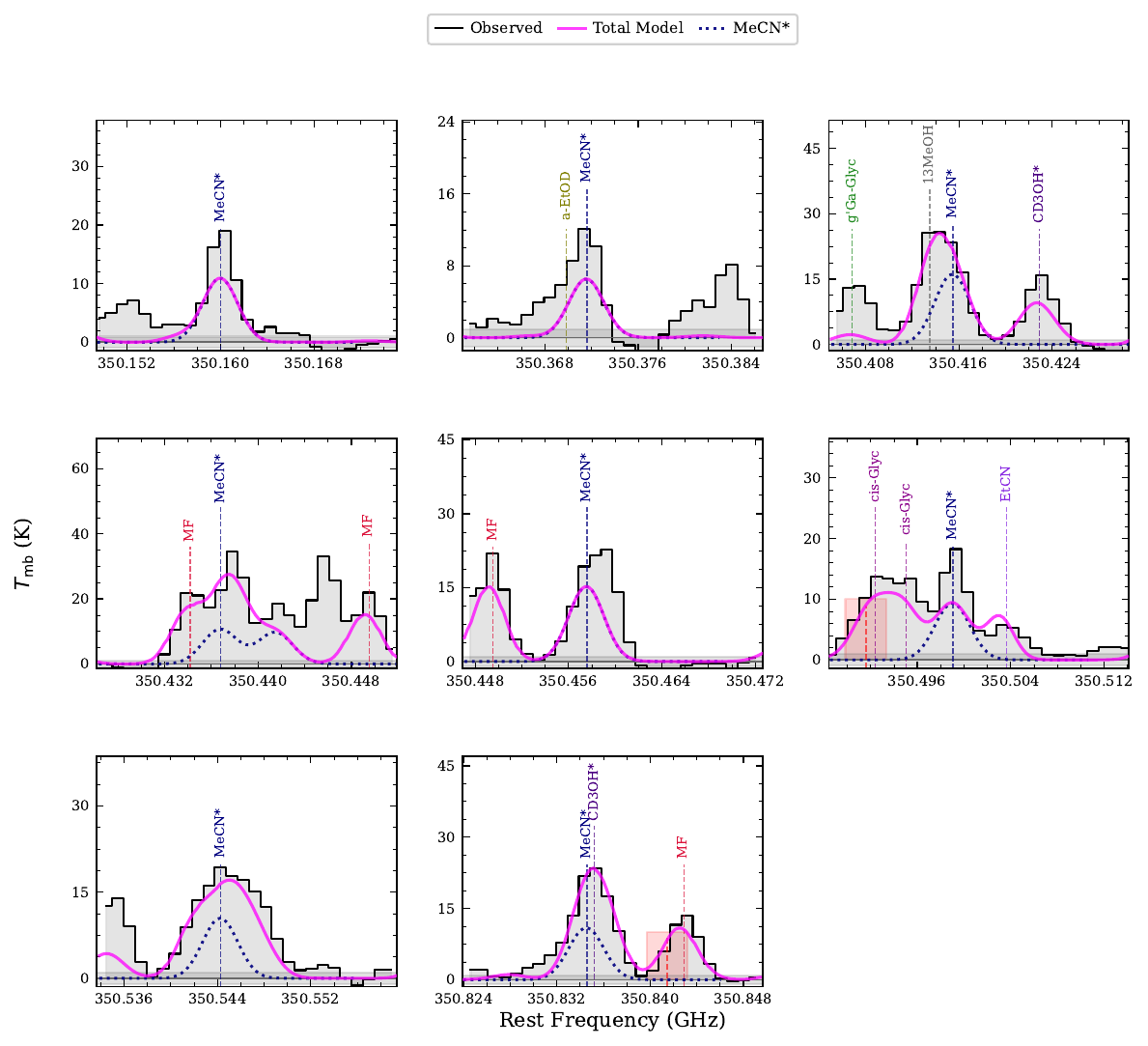}
    \caption{Synthetic LTE spectra of Vibrationally Excited Methyl Cyanide ($CH_3CN, v_8=1$). Same as Fig.~\ref{fig:ch3ocho} but for \ch{CH_3CN (v_8 = 1)}}
    \label{fig:ch3cn}
\end{figure*}

\clearpage

\bibliography{achemso-demo}
\bibliographystyle{aasjournal}

\end{document}